\documentstyle[preprint,aps]{revtex}

\tightenlines

\begin{document}

\draft

\title{\bf Studies of one- and two-hole states\\
in the 2D $t\hbox{-}J$ model  via series expansions
}
\author{C.J. Hamer\cite{byline1}, Zheng Weihong\cite{byline2}, and J. Oitmaa\cite{byline3}}
\address{School of Physics,
The University of New South Wales,
Sydney, NSW 2052, Australia.}

\date{June 26, 1998}

\maketitle

\begin{abstract}
We study one and two hole properties of the $t\hbox{-}J$ model at half-filling
on the square lattice using series expansion methods at $T=0$.
The dispersion curve for one hole excitations is
calculated and found to be 
qualitatively similar to that obtained by other methods,
but the bandwidth for small $t/J$ is some 20\% larger than
given previously. We also obtain the binding energy and  dispersion relation
for two hole bound states. The lowest bound state as $t/J$ increases
is found to be first $d$-wave, and then $p$-wave, in accordance
with predictions based upon the Kohn-Luttinger effect.
We also make a similar study for the $t\hbox{-}J_z$ model.
\end{abstract}


\narrowtext
\section{INTRODUCTION}
Models of strongly correlated electrons in two dimensions
have been much studied in recent years as
part of a search for a minimal model of the high-T$_c$ 
superconducting cuprates, and other  fascinating
new antiferromagnetic systems\cite{dag94,dag96,Lu93}. The
$t\hbox{-}J$ model, defined by the Hamiltonian
\begin{equation}
H=-t \sum_{\langle ij\rangle, \sigma} P (c_{i,\sigma}^{\dag} c_{j,\sigma} 
   + c_{j,\sigma}^{\dag} c_{i,\sigma} ) P +
 J \sum_{\langle ij\rangle } ({\bf S}_i \cdot {\bf S}_j - {1\over 4} n_i n_j )
\end{equation}
has become a generic model for mobile electrons/holes in doped 
antiferromagnets. In this expression the $c_{i,\sigma}^{\dag}$,
$c_{i,\sigma}$ are the usual electron creation and
destruction operators, ${\bf S}_i$ are the electron spin operators and
the sum is over nearest neighbors pairs. $P$ is a projection
operator which ensures that doubly occupied states are
excluded.

The $t\hbox{-}J$ model originally was derived\cite{hir85} as
the strong correlation limit of the Hubbard model, with $J=2t^2/U$,
but it is now generally treated as an effective Hamiltonian
in its own right, with $t$ and $J$ independent parameters.
For the cuprates it is estimated\cite{hyb90} that $t/J\simeq 3$,
which is well outside the region of validity of the Hubbard
mapping. An independent ``derivation'' of the $t\hbox{-}J$ model as
an effective Hamiltonian for the cuprates has been presented by
Zhang and Rice\cite{zha88}.

At exactly half-filling (one electron per site) the kinetic
term vanishes because of the single-occupancy restriction
and the model reduces to an antiferromagnetic insulator.
It is known from a variety of numerical and analytic 
calculations\cite{man91} that this model, on the square
lattice, has nonzero long range antiferromagnetic (N\'eel)
order, reduced by strong quantum fluctuations. Removal of a
small number of electrons (by doping) will allow
mobility of holes and will reduce the antiferromagnetic
order. The physics of a small number of holes in a dynamic
antiferromagnetic background remains a challenging problem.
Previous studies of the $t\hbox{-}J$ model at $T=0$ have used a variety
of approaches. Exact diagonalization methods have been used extensively
on clusters up to 32 sites\cite{bar89,poi94,leu95,lee97}, but suffer from
substantial finite-size corrections and an inability to treat 
extended excitations. Green's function Monte Carlo calculations have
been performed for larger lattices, up to $8\times 8$\cite{bon},
and show, for example, a significant decrease of the two-hole binding energy 
with increasing lattice size. A variety of variational approaches and analytic
many body methods have also been employed\cite{liu92,ede92,yu95,li96,bel97,shr90,kuc93}.

The picture that emerges from this body of work is roughly as follows.
For one hole the ground state energy is given by
\begin{equation}
\epsilon_{1h}/t \simeq -3.2 + 2.8 (J/t)^{0.73}
\end{equation}
where the coefficients are approximate numerical estimates. The dispersion
curve for 1-hole states is qualitatively similar from different methods, with
minima at $(\pi/2,\pi/2)$
and equivalent  points. The bandwidth from the largest diagonalizations scales as
\begin{equation}
W/t \simeq -0.33 + 2.04 (J/t)^{0.64}
\end{equation}
where the coefficients are again estimates and are not known with great
precision. The spectral function for 1-hole states has been 
calculated\cite{leu95} and shows a clear quasiparticle peak,
together with a continuum at higher energies.
For two holes the binding energy, spectral function and
pair susceptibility have been computed. The results 
support the existence of a bound  state with 
$S=0$ and d-wave symmetry for $J> J_c$, with $J_c \simeq 0.3t$
although not all methods agree and this is still controversial.
The dispersion curve for two-hole bound states is also of interest
$-$ to our knowledge the only calculation of this is due to
Eder\cite{ede92}.

The aim of the present work is to study the physics of one-hole
and two-hole states in the half-filled 
$t\hbox{-}J$ model at zero temperature via linked-cluster
series expansions. This approach has been developed extensively
in recent years by our group\cite{he90,oit94,oit96} and others\cite{sin89,gel90,gelfand}.
Most of the applications to date have been to spin systems, although we have included fermions in
lattice gauge calculations. Shi and Singh\cite{shi95} have used this
approach to study the Hubbard model at half filling, via an expansion about an
additional Ising term included in the Hamiltonian. For completeness
we also mention the use of high-temperature expansions to study the 
$t\hbox{-}J$ model\cite{put92,sin92}. The present work of course addresses different
questions which cannot be probed by high-temperature series.

The plan of the paper is as follows.  In Sec. II, the series expansion method
is briefly reviewed, and the results for one-hole properties of the $t\hbox{-}J$ model are
presented. In Sec. III, we study the dispersion relation
and the binding energy for two-hole bound states.
In Sec. IV, we study  the one-hole and two-hole  properties of the $t\hbox{-}J_z$ model. Sec. V is
devoted to a summary and discussion.

\section{One-Hole Properties}

The series expansion method is based on a linked cluster
formalism associated with standard Rayleigh-Schr\"odinger perturbation theory.

The ground state energy $E_0^N$ for a lattice of $N$ sites can be
written as a sum over all connected clusters $\{\alpha\}$ as
\begin{equation}
E_0^N = \sum_{\alpha} C_{\alpha}^N \epsilon_{\alpha} \label{eq2}
\end{equation}
where $C_{\alpha}^N$ is a geometrical embedding factor and
$\epsilon_{\alpha}$ is the ``cumulant energy'' of cluster $\alpha$.
The same formula can be used for the energy of any particular cluster $\beta$
\begin{equation}
E_0^{\beta} = \sum_{\alpha} C_{\alpha}^{\beta} \epsilon_{\alpha} \label{eq3}
\end{equation}
where the sum is now over all sub-clusters embeddable in $\beta$.
The cluster energy is calculated perturbatively, then the cumulant energies
are obtained iteratively from Eq. (\ref{eq3}) and substituted into 
Eq. (\ref{eq2}) to obtain the final series. The formalism for excited
states, needed to compute the dispersion curves, is a little
more complex\cite{gelfand}.

We introduce an Ising spin anisotropy
\begin{eqnarray}
J ( {\bf S}_i \cdot {\bf S}_j -{1\over 4} n_i n_j) &\to  &
J_z ( S_i^z S_j^z -{1\over 4} n_i n_j ) 
+ J_{xy} (S_i^x S_j^x + S_i^y S_j^y ) \nonumber \\
&=& J_z [ ( S_i^z S_j^z  -{1\over 4} n_i n_j ) + x (S_i^x S_j^x + S_i^y S_j^y )]
\end{eqnarray}
where $x=J_{xy}/J_z$, and rewrite the Hamiltonian for the $t\hbox{-}J$ model as
\begin{equation}
H/J_z = H_0 + x V
\end{equation}
with
\begin{eqnarray}
H_0 &=& \sum_{\langle ij\rangle}
 (S_i^z S_j^z - {1\over 4} n_i n_j ) + r \sum_i (-1)^i S_i^z  \\ 
V &=& - {t\over J_{xy}} \sum_{\langle ij\rangle ,\sigma}
P (c^{\dag}_{i,\sigma} c_{j,\sigma} +
c^{\dag}_{j,\sigma} c_{i,\sigma} ) P \nonumber \\
&& + \sum_{\langle ij\rangle} (S_i^x S_j^x + S_i^y S_j^y)
- r \sum_i (-1)^i S_i^z 
\end{eqnarray}
with a staggered $z$-field of strength $r$ to improve convergence\cite{shi95}.
This term vanishes in the total Hamiltonian in the isotropic
exchange limit $x\to 1$. 
The Ising part of the exchange, $H_0$, forms
the unperturbed Hamiltonian while the perturbation $V$ includes both
the hole hopping terms and transverse spin fluctuations.
The original Hamiltonian for the $t\hbox{-}J$ model is recovered 
in the limit $x=1$ (i.e. $J=J_z=J_{xy}$). 
We expect the series expansion to give more reliable results in the
small $t/J$ region, where the perturbation $V$ is less important.

We directly compute the series for the ground state energy difference between
the system with one (or two) hole(s) and the system with no hole.
Two types of series can be obtained in this way. By fixing 
$y\equiv t/J_{xy}$ and $r$
we can compute series in the single variable $x$ up to order $L$
(where $L$ is typically of order 11-13),
\begin{equation}
E/J_z = \sum_{i=0}^L c_i(y,r) x^i
\end{equation}
We refer to these as $x$-series. Extrapolation to $x=1$ is carried out by
standard Pad\'e approximant or integrated differential approximant methods\cite{gut}.
In the other approach we 
keep both $x$ and $t/J_z$ as 
expansion parameters and obtain double series of the form
\begin{equation}
E/J_z = \sum_{ij} d_{ij}(r) (t/J_z)^i x^j
\end{equation}
where the coefficients $d_{ij}$ are computed up to order
$i+j\le L$. Extrapolation to $x=1$
then yields a series in $t/J$, 
\begin{equation}
E/J = \sum_i e_i (t/J)^i
\end{equation}
which we refer to as a $t$-series.

We first consider states with one hole,
which may exist either on the up-sublattice (A) or
the down-sublattice (B), where the up and down sublattice are  
represented in Fig. \ref{fig_lattice}(a) by open and full circles, respectively.
Here $(e_x,e_y)$ are the usual primitive translation vectors.
The unperturbed wave function for the zero-momentum spin-up state is taken to be
\begin{equation}
\vert \Psi \rangle = \sum_{i\in A} c_{i,\uparrow} 
\vert N\rangle 
\end{equation}
where $\vert N\rangle$ is the classical Ne\'el state.
Now a hole in the up (down) sublattice can never transform into
a single hole in the down (up) sublattice, since the Hamiltonian
conserves the total spin $\sum_i S_i^z$; so that we should choose 
$(e_{x'}, e_{y'})$ shown in Fig. \ref{fig_lattice}(a)
as the primitive translation vectors, rather than $(e_x, e_y)$.
The unit vectors $(e_{x'}, e_{y'})$ are those of the antiferromagnetic
bipartite lattice. The Brillouin zones in reciprocal space
corresponding to these different choices are shown in Fig. \ref{fig_lattice}(b).
Most previous calculations, however, have used $(e_x,e_y)$
as unit vectors (corresponding to $(k_x,k_y)$ in momentum space),
instead of  $(e_{x'},e_{y'})$, so we will also conform to this
convention (and set $a$ to one) unless it is specified otherwise.

For momentum ${\bf k}=0$ we obtain the energy to order $L=13$, involving
8739 distinct clusters containing up to
13 sites. For ${\bf k}\neq 0$ it is necessary to distinguish 
between clusters having different spatial orientation. We expand
the energy to order $L=11$, involving 23546 clusters with up to
11 sites. 
The series coefficients for $y=0.5$ at fixed momentum $(k_x,k_y)=(\pi/2,\pi/2)$ 
and different $r$ are listed in Table \ref{tab_1hmin}, where we can see
the series converges best around $r=2$.
The extrapolations of these series by 
using the integrated differential approximant methods
are shown in Figure \ref{fig_diffr}, where we can see that at $x=1$,
the extrapolations from different $r$ series agree with each other very well, 
while the case $r\simeq 2$ gives the smallest error bar.
In Table \ref{tab_tjK}, we list the full spectrum series for $y=0.5$ and $r=0$.
The other coefficients are too lengthy to publish but can be supplied on
request. 

In Figure \ref{fig_Ek1h} we show the energy of a one-hole state for 
${\bf k}$ along symmetry directions in the Brillouin 
zone, for various ratios  $t/J$ obtained from $x$-series. 
The lowest energy occurs at $(k_x,k_y)=(\pi/2, \pi/2)$ and
equivalent points. This result, and the overall shape of the one-hole
dispersion curve are in very good agreement with previous work\cite{liu92,ede90,mar91,leu95,lee97}.
We note that the symmetry and the redundant information evident in
Fig. \ref{fig_Ek1h} are the results of choosing a Brillouin
zone which is two times too big. If we choose 
$(k_{x'}, k_{y'})$ as defining the bipartite Brillouin zone,
the corresponding dispersion relation along symmetry directions
in the Brillouin zone is shown in Fig.  \ref{fig_Ek1h2}.
Figure \ref{fig_W} shows the one-hole bandwidth, $W=E(0,0)-E(\pi/2,\pi/2)$,
as a function of
$J/t$, compared with results from other approaches.
For $J/t$ around 0.5 our results agree well with other work,
but for large $J/t$ our result is  significantly larger 
than that obtained by Martinez and Horsch\cite{mar91}.
Also we found a peak value of $W/t \simeq 1.15$ at $J/t \simeq 1.25$,
rather than $W/t\simeq 1$ at $J/t=0.8$ as obtained by 
Martinez and Horsch\cite{mar91}, and Liu and Manousakis\cite{liu92}.

Using the two-variable expansion we obtain the one-hole energy $E({\bf k})$
at some particular values of ${\bf k}$:
\begin{eqnarray}
E^{\rm 1h} (0,0)/J &=& 2.164(2) -0.655(8) (t/J)^2 - 0.031(5) (t/J)^4 + O[(t/J)^6] \nonumber \\
E^{\rm 1h}(\pi/2,\pi/2)/J &=& 2.164(2) -3.91(4) (t/J)^2 + 9.1(4) (t/J)^4 + O[(t/J)^6] \label{eq_e1h} \\
E^{\rm 1h}(\pi,0)/J &=& 2.164(2) -2.68(2) (t/J)^2 + 3.06(4) (t/J)^4 + O[(t/J)^6]\nonumber
\end{eqnarray}
These coefficients are not exact as they involve an extrapolation to $x=1$. 
The leading coefficient 2.164(3) can be compared with 2.1552 from 
a spin wave calculation\cite{bul89} and
2.193(7) from a Monte Carlo simulation\cite{bar90}.

The bandwidth can be obtained as
\begin{equation}
W/J= 3.26(5) (t/J)^2 -9.1(4) (t/J)^4 + O[(t/J)^6] \label{W_2p}
\end{equation}
This result is also plotted in Figure \ref{fig_W} as a dashed line,
and it agrees well at small $t/J$  with that obtained from $x$-series.
The leading coefficient 3.26(5) should be compared with previous
calculations giving a value of 2\cite{mar91} or 2.6\cite{sch89}.

\section{Two-hole properties}
In the previous section we developed a series expansion approach to calculate
one-hole properties of the $t\hbox{-}J$ model, and showed how reliable quantitative
results could be obtained. In the present section we extend this
approach to consider states with two holes.

We start from an unperturbed wave-function consisting of two
nearest-neighbour holes with opposite spin, and overall momentum 
${\bf k}$:
\begin{equation}
\vert \Psi ({\bf k}) \rangle = \sum_{i\in A} c_{i,\uparrow} e^{i {\bf k}\cdot {\bf x}_i}
( c_{i-{e_x},\downarrow} e^{-i k_x a/2} 
+ c_{i+{e_x},\downarrow} e^{ i k_x a/2}
+ c_{i-{e_y},\downarrow} e^{-i k_y a/2}
+ c_{i+{e_y},\downarrow} e^{ i k_y a/2}
) \vert {\rm N} \rangle 
\end{equation}
Here again there are two different choices of Brillouin zone possible. One may choose 
to use a diagonal unit vector of length $a/\sqrt{2}$ (where $a$ is the lattice spacing), 
which connects the centre points of two adjacent bonds of the original lattice
as shown in Fig. \ref{fig_lattice2h}(a); this corresponds to the large, outer 
Brillouin zone in Fig. \ref{fig_lattice2h}(b). This was basically the convention
used by Eder\cite{ede92}. Alternatively, and more correctly, one
may choose a diagonal unit vector of length $\sqrt{2}a$, differentiating between
the $A$ and $B$ sublattices. This gives rise again to the Brillouin
zone of the bipartite lattice, marked as the inner zone in Fig.
\ref{fig_lattice2h}(b).
In this second case, there are 4 independent bond states in the unit
cell, and one may form linear combinations of them making up the
conventional $s$-, $p_x$-, $p_y$- and $d$-wave states. With the first choice,
all these four states are included within one Brillouin zone:
the points corresponding to zero-momentum $s$-, $p_x$-, $p_y$- and $d$-wave
states are shown\footnote{The ``$p_x$-'' and ``$p_y$-'' wave states are
degenerate by reflection symmetry. The points marked $p_x$ and $p_y$
in Fig. \ref{fig_lattice2h}(b) are really linear combinations of the
proper $p_x$ and $p_y$ states.} in Fig. \ref{fig_lattice2h}(b). Then one 
finds that the different states are related by equations such as
\begin{equation}
E^{2d}(k_x,k_y) = E^{2s} ( \pi-k_x, \pi-k_y )
\end{equation}
(in units where $a/\sqrt{2}$ has been rescaled to one). This relationship is 
evident in Fig. 19 of Ref. \cite{ede92}.

We have computed series for the two-hole spectrum $E^{\rm 2h}({\bf k})$
 up to order $L=11$. This
involves 46,440 distinct clusters of up to 12 sites. The $x$-series coefficients for $y=0.5$ and
$r=0$ are listed in Table \ref{tab_tjK}. 
 
Using the two-variable expansion, after extrapolation to $x=1$, we obtain
the following estimates for the two-hole $s$-, $p$- and $d$-wave pair energy:
\begin{eqnarray} 
E^{2d}/J &\equiv & E^{\rm 2h}(0,0)/J =     3.798(2) - 7.66(3) (t/J)^2 + 17.5(5) (t/J)^4 + O[(t/J)^6]\nonumber \\
E^{2p}/J  &\equiv &E^{\rm 2h}(0,\pi)/J =   3.798(2) - 3.705(10) (t/J)^2 - 4.1(1) (t/J)^4 + O[(t/J)^6]  \label{eq_e2h} \\
E^{2s}/J  &\equiv &E^{\rm 2h}(\pi,\pi)/J = 3.798(2) - 0.52(3) (t/J)^2 - 1.62(5) (t/J)^4 + O[(t/J)^6] \nonumber 
\end{eqnarray}


So 
 the binding energies defined by
\begin{equation}
E_{\rm b} ({\bf k}) \equiv E^{\rm 2h} ({\bf k}) - 2 E^{\rm 1h} (\pi/2,\pi/2)
\end{equation}
for $s$-, $p$- and $d$-wave pair states are:

\begin{eqnarray}
E_{\rm b}^{2d}/J &=& -0.529(4) + 0.16(10) (t/J)^2 - 0.7(13) (t/J)^4 
      + O[(t/J)^6] \nonumber \\
E_{\rm b}^{2p}/J &=& -0.529(4) + 4.1(1) (t/J)^2 - 22.3(9) (t/J)^4
      + O[(t/J)^6] \label{eq_binding} \\
E_{\rm b}^{2s}/J &=& -0.529(4) + 7.3(1) (t/J)^2 - 19.8(9) (t/J)^4 
      + O[(t/J)^6] \nonumber
\end{eqnarray}


The binding energy for $t=0$ has previously been given as
$-0.51$ by spin wave theory\cite{bul89}, and $-0.58(2)$ by Monte Carlo\cite{bar90}.
Our result is in relatively good agreement but suggests that the
Monte Carlo calculations may overestimate the binding energy
somewhat.

We have also computed the dispersion curves for the two-hole
bound state from $x$-series, and these are shown in Figure \ref{fig_Ek2h}, 
for various small $t/J$.
The minimum occurs at $(\pi,\pi)$ ($d$-wave pair), and the maximum
occurs at $(0,0)$ ($s$-wave pair), while the energy at $(0,0)$ is almost unchanged
as $t/J$ increases.
The curves are quite flat for small
$t/J$, as expected, and develop more structure as $t/J$ increases.
The results  are broadly similar to those of Eder\cite{ede92}.

The $s$-, $p$- and $d$-wave pair binding energies, obtained 
from both the $x$-series and the two-variable expansion as a function of $t/J$,
are shown in Figure \ref{fig_binding}. Note that for $x$-series,  the extrapolation is
done on series for $E_{\rm b} ({\bf k})$, rather than 
the 1-hole and 2-hole energies individually, because values of these
two energies are quite large, while their difference is usually small.
The convergence 
is excellent for small $t/J$,
and we believe our results are very accurate there. The uncertainty
increases  for larger $t$.
We can see from this figure that as $t/J$ increases from zero,
the $s$-wave binding energy decreases very quickly, and the binding
 disappears around $t/J\simeq 0.295$.
The $p$-wave binding energy also decreases quickly in the small $t/J$ region,
but it turns out to be very flat beyond  $t/J\sim 1$, and the binding
energy is very small.
The $d$-wave binding energy decreases slightly as $t/J$ increases from zero. 
Our results are consistent with
the $d$-wave binding disappearing around $t/J\sim 2.5$.
Beyond that point, the $p$-wave state appears to have
the lower energy, and remains very weakly bound according
to our results.

Figure \ref{fig_binding} also shows some other numerical results for the
2-hole binding energy. The solid circles with error bars are GFMC results for
the $8\times 8$ lattice obtained by Boninsegni and  Manousakis\cite{bon},
and lie somewhat below our result, though broadly similar in
trend: the discrepancy may easily be attributed to a finite-size effect.
The solid triangles are the  results of 
a recent 32-site finite-lattice calculation\cite{che98}.
The open circles are the results of 
a linear extrapolation in $1/N$ from the finite-lattice data by
Chernyshev,  Leung and Gooding\cite{che98};
but they discuss reasons why such a naive
extrapolation is not to be relied on especially at large $t/J$.
The results lie somewhat above our series results
for the $d$-wave binding energy, but again have a broadly
similar trend. Chernyshev,  Leung and Gooding\cite{che98} find  that 
the lowest bound state on the 32-site lattice is a singlet
state with $d_{x^2-y^2}$ symmetry.


The apparent cross-over between the $p$-wave and $d$-wave bound states came
initially as a surprise. But in fact it accords very well with
predictions\cite{bar92,chu93} based on the Kohn-Luttinger effect.
In a repulsive Fermi liquid, pairing in high-angular momentum states
can occur due to a singularity in the 2-particle scattering amplitude.
This effect was previously believed not to occur in two dimensions,
because it does not appear up to second order in perturbation theory;
but it was recently shown\cite{bar92,chu93} that a singularity
does arise at higher order. Chubukov\cite{chu93} showed that in the
dilute limit the effect survives down to $l=1$ and gives rise to $p$-wave
pairing. He also pointed out that the same effect will occur in an
arbitrarily dense Fermi liquid. Both numerical\cite{dag93}
and analytical\cite{hel97,kag94} approaches have shown that at
low densities $p$-wave pairing is dominant for
$J/t \ll 1$, $d$-wave is dominant at $J/t > 1$, and there is a
region of $s$-wave dominance for $J/t \gtrsim 2$.
Figure \ref{fig_binding} appears to indicate that a similar pattern extends
even to the region of half-filling.
In fact, the earlier variational calculation\footnote{There is a discrepancy
between Eder's results and ours, which is due to his omission of the
term $-1/4 n_i n_j$ in the Hamiltonian.} of Eder also shows a crossover
between $p$-wave and $d$-wave states (Fig. 21 of Ref. \cite{ede92})
at around $t/J=2$.

Several theoretical approaches\cite{bel97,shr90,kuc93}
have predicted that magnon exchange should give
rise to shallow, long-range bound states even at
large $t/J$ values. The behaviour of the $p$-wave
state looks consistent with this idea, but not
the $d$-wave state. It is possible that another state (or states)
in the $d$-wave sector could cross over the one we
have calculated, and become the lowest energy 
state at large $t/J$. Such a crossover could typically
imply a square-root singularity in the energy eigenvalue,
and the series extrapolation would no longer be valid
beyond that point.

\section{ \hbox{t-J$_z$} model}
We also study a simplified version of the $t\hbox{-}J$ model,
the $t\hbox{-}J_z$ model,  which
has an Ising rather than a Heisenberg spin interaction
\begin{equation}
H = J_z \sum_{\langle ij\rangle}
 (S_i^z S_j^z - {1\over 4} n_i n_j ) 
- t \sum_{\langle ij\rangle ,\sigma}
P (c^{\dag}_{i,\sigma} c_{j,\sigma} +
c^{\dag}_{j,\sigma} c_{i,\sigma} ) P \label{H_tJz}
\end{equation}
Here we naturally take the first term in $H$ as the unperturbed
Hamiltonian, and the second term as the perturbation, and 
perform an expansion in $t/J_z$. This expansion
only converges well up to about $t/J_z\sim 2$. To improve
the convergence  in the large $t/J_z$ region, we can include a
 staggered $z$-field of strength $r$ as in the $t\mbox{-}J$ model, and
perform the following separation,
\begin{equation}
H/J_z = H_0 + x V \label{H_tJz2}
\end{equation}
where
\begin{eqnarray}
H_0 &=& \sum_{\langle ij\rangle} (S_i^z S_j^z - {1\over 4} n_i n_j ) 
 + r \sum_i (-1)^i S_i^z  \\ 
V &=& - {t\over J_{z}} \sum_{\langle ij\rangle ,\sigma}
P (c^{\dag}_{i,\sigma} c_{j,\sigma} +
c^{\dag}_{j,\sigma} c_{i,\sigma} ) P - r \sum_i (-1)^i S_i^z 
\end{eqnarray}
We then take $H_0$ as the unperturbed Hamiltonian and $V$ as the perturbation, and 
perform an expansion in $x$ for fixed values of $t/J_z$ and $r$. 
Here again we need to extrapolate
the series to $x=1$.

The series for the one-hole state have been computed up to order 21 (that is,
to $(t/J_z)^{21}$
for the Hamiltonian in Eq. (\ref{H_tJz}) or to order $x^{21}$ for the
Hamiltonian in Eq. (\ref{H_tJz2}))
involving a list of 23546 clusters with up to 11 sites,
while the series for the two-hole bound state have
been computed up to order 18,
involving a list of 23546 clusters up to 11 sites.
The final series in $t/J_z$ for the spectrum of 
the one-hole state and two-hole bound state 
are listed in Table \ref{tabtJz}.

Fig. \ref{fig_Ek1h_tJz} shows the one-hole spectrum for 
some particular values of $t/J_z$.
Here we can see that the shape of the dispersion curve is quite 
different from that for the $t\hbox{-}J$ model:
the lowest energy occurs at $(0,0)$ rather than $(\pi/2,\pi/2)$
as in the $t\hbox{-}J$ model, and the bandwidth here is about 10 times
smaller than that for the $t\hbox{-}J$ model. 
The minimum one-hole energy as a function of $J_z/t$ is shown in 
Fig. \ref{fig_E1hmin}. In the interval $0.2 \leq J_z/t \leq 1$,
the minimum one-hole energy can be very well fitted\cite{rie93,bar89} as
\begin{equation}
E^{1h}/t = -3.60 + 2.9 (J_z/t)^{2/3} + J_z/t 
\end{equation}
where  the linear term comes from the 
$-1/4 n_i n_j$ term in the Hamiltonian. This is in good agreement
with previous estimates\cite{rie93,bar89}.


The two-hole dispersion is shown in Fig. \ref{fig_Ek2s_tJz}.
We note that the energy at $(0,0)$ is exactly $7J_z/2$, independent of $t$,
and the energy along path $(\pi,0)$ ($p$-wave pair) to $(\pi,\pi)$
($d$-wave pair) is very flat.

The $s$-, $p$- and $d$-wave binding energies defined by
\begin{equation}
E_{\rm b}({\bf k}) \equiv  E^{2h}({\bf k})-2 E^{1h}({\bf k}=0)
\end{equation}
are shown in Figure \ref{fig_binding_tJz}.
We can see from this figure that $s$-wave binding disappears around
$t/J_z\simeq 0.32$. The $d$-wave  binding energies are slightly greater
than for $p$-wave when $t/J_z \lesssim 0.83$, but when $t/J_z \gtrsim 0.83$ , 
the $p$-wave pair has a greater binding energy than the $d$-wave, similarly
to that found for the full $t\hbox{-}J$ model. Thus the $t\hbox{-}J_z$ model also shows a crossover
between  $d$-wave and $p$-wave bound states, occurring even earlier than in the
full $t\hbox{-}J$ model.
Again, the earlier variational calculations of Eder showed 
very similar results (Fig. 12 of Ref. \cite{ede92}).

The solid points in Figure \ref{fig_binding_tJz} are the results of
a reduced basis diagonalization study
on clusters of up to 50 sites by Riera and Dagotto\cite{rie93}.
 The trend of these results
agrees extremely well with the series estimates, although their 
estimates lie systematically a little lower. They
do not discuss the symmetry of the bound state\cite{rie93}. They find the binding
of the 2-hole state disappears at $J_z/t\simeq 0.183$, or
$t/J_z\simeq 5.5$, which agrees very well with our results.




\section{CONCLUSIONS}
The $t\hbox{-}J$ model on the square lattice has been studied by many 
workers over the last decade, using a variety of techniques.
These methods are all approximate and have uncertain accuracy.
Independent confirmation from a variety of approaches
increases one's confidence  in the results.

We have used the linked-cluster series expansion method
to study one and two hole properties of 
 the half-filled $t\hbox{-}J$ and $t\hbox{-}J_z$ models in 2-dimensions at $T=0$.
To our knowledge this is the first time this has been done.
Our results for 1-hole and 2-hole states can be summarized as follows:

\begin{itemize}
\item The 1-hole dispersion curve has been computed 
throughout the Brillouin zone, and agrees in shape with 
previous calculations.

\item The minimum for 1-hole dispersion occurs at $(\pi/2, \pi/2 )$ and equivalent
points, again in agreement with previous work.

\item The bandwidth $W$ for the 1-hole dispersion curve of the $t\hbox{-}J$ model
has been calculated. At small $J/t$
it agrees with the spin-wave calculations of Martinez and  Horsch\cite{mar91}
and Liu and  Manousakis\cite{liu92}, but beyond $J/t \simeq 1$ our
results lie substantially higher than theirs. We get a peak value of 
$W/t\simeq 1.15$ at $J/t\simeq 1.25$, rather than  $W/t\simeq 1$ 
at $J/t\simeq 0.8$ as given 
in previous work.

\item We have computed the dispersion curve for
2-hole bound states. To our knowledge this has 
not been well calculated previously.

\item In the limit $t/J=0$ of the $t\hbox{-}J$ model,
we can compare our results for the one-hole 
and two-hole energy ($E^{1h}$ and $E^{2h}$), 
the binding energy ($E_{\rm b}$), and the bandwidth
for the one-hole state ($W$)
with previous spin-wave\cite{bul89},
Monte Carlo\cite{bar90} and other calculations.


\begin{tabular}{l|cl}
              &This Work~~~~~~ &  Previous works  \\
\hline
$E^{1h}/J$~~~ &  2.164(2)  & 2.155(SW), 2.193(MC) \\
$E^{2h}/J$    &  3.798(2)  & 3.79436(SW), 3.801(8)(MC) \\
$E_{\rm b}/J$ &  -0.529(4) &  -0.51(SW), -0.58(2)(MC) \\
$W$           &  $3.26(t/J)^2$ &  $2(t/J)^2$\cite{mar91},~$2.6(t/J)^2$\cite{sch89} \\
\end{tabular}

We can see that our results agree very well with other results,
except for $W$, where we find a significantly larger value for 
the leading coefficient.

\item At large $t/J$, we find a crossover in binding energy between
the $d$-wave and $p$-wave 2-hole states, for both the $t\hbox{-}J$
and $t\hbox{-}J_z$ models. 
This phenomenon 
is likely to arouse some controversy:
it certainly needs to be checked by other methods. As supporting
evidence, we note the following points:

i) The trend of the series results for the $t\hbox{-}J_z$ model agrees very well 
with numerical calculations of Riera and Dagotto\cite{rie93};


ii) The phenomenon would be in accord with
the Kohn-Luttinger effect as predicted by Baranov, Kagan and Chubukov\cite{bar92,chu93},
and confirmed at low densities\cite{hel97,dag93,kag94}.
Chubukov pointed out that the same effect should occur in an arbitrarily
dense Fermi liquid.

iii) Very much the same phenomenon was observed previously in the
variational calculations of Eder\cite{ede92}. The concordance
between his results and ours would seem to indicate that
his variational ansatz is a good one.

iv) A similar enhancement of pairing in the odd-parity singlet
channel, analogous to our $p$-wave state, was also noticed for
the Hubbard model by Scalettar, Singh and Zhang\cite{sca91}.

Another possibility, however, is that another state in the 
$d$-wave sector may cross over the one we have tracked, and
become the lowest energy state at large $t/J$. Such a phenomenon
would not be detected by our series analysis.



\end{itemize}

The formation of a two-hole bound state is not of itself 
sufficient to prove the existence of either a superconducting
condensate, or of ``phase separation" between filled
and empty sites\cite{dag94}; but one would certainly
presume it to be a necessary precondition for either of these
phenomena to occur.
Our results indicate that the $d$-wave binding energy vanishes
at around $t/J\simeq 2.5$. The $p$-wave binding energy
may continue a little longer, but it is certainly very weak, and 
our error bars are not small enough to allow any definite conclusion on
this point. Boninsegni and  Manousakis\cite{bon}
and Poilblanc {\it et al.}\cite{poi94}
found the binding to disappear at around $t/J\simeq 3.7$.

The question of phase separation is still under debate: see the review
by Dagotto\cite{dag94}. At small $t/J$ phase separation is known to occur.
Emery, Kivelson and Lin\cite{eme90} suggested that in fact phase
separation occurs at {\it all} couplings in the neighborhood of 
half-filling. Subsequently, the majority of studies\cite{dag94,hel97}
concluded that at intermediate coupling ($0.3<J/t<1$) there is 
a region where binding of pairs occurs, but no phase separation.
Poiblanc\cite{poi95} and Kohno\cite{koh97} put the phase
separation boundary at about $t/J\simeq 1.3$. But Hellberg
and Manousakis\cite{hel97} have recently challenged this scenario,
arguing that the criterion of a vanishing inverse compressibility used in previous 
studies was unreliable. They use a Maxwell construction technique,
from which they conclude that indeed phase separation occurs
at all couplings in accordance with the original hypothesis
of Emery {\it et al.}\cite{eme90}. Our results unfortunately have no
direct bearing on this debate. To explore these questions one must treat
the model away from half-filling, which is not possible using
our present series technique, because it cannot handle an unperturbed
ground state which is ``doped" away from half-filling, and
is therefore inhomogeneous.

The presence or absence of a superconducting condensate is also a matter of
debate\cite{dag94}. One recent study\cite{shih98} found no evidence
of $d$-wave superconducting correlations in the {\it physical}
parameter range $t/J\ge 2$. The simple $t\hbox{-}J$ model is beginning to seem a little
doubtful as a theory of the high-T$_c$ superconductors.

There are a number of other significant physical
quantities, such as spectral functions,  spectral
weights, and magnetic susceptibilities which are the
subject of ongoing work.

\acknowledgments
We are grateful to Profs. H.Q. Lin and O. Sushkov  for  their useful advice,
and to Profs. Rajiv R.P. Singh, R. Eder and A.L. Chernyshev for very helpful
 correspondence.
This work forms part of a research project supported by a grant
from the Australian Research Council. The computation has been performed
on Silicon Graphics Power Challenge and Convex machines. We thank the New
South Wales Centre for Parallel Computing for facilities and assistance
with the calculations.


\begin{figure} 
\caption{(a) The open (full) points represent the up (down) sublattices, respectively.
The primitive translation vectors, $(e_x, e_y)$ and $(e_{x'}, e_{y'})$, are shown
by the arrows. (b) The reciprocal space of these lattices. The Brillouin zones for
$(e_x, e_y)$  and $(e_{x'}, e_{y'})$ are $-\pi/a< k_x,k_y\leq \pi/a$ 
and $-\pi/(\protect\sqrt{2}a)< k_{x'},k_{y'}\leq \pi/(\protect\sqrt{2}a)$, respectively,
where $a$ is the lattice spacing.
}
\label{fig_lattice}
\end{figure}

\begin{figure} 
\caption{The extrapolation of the $x$-series for 
one-hole energy $E^{1h}/J_z$ for $y=0.5$ at momentum ${\bf k}=(\pi/2,\pi/2)$
and several different $r$ values.
Several different integrated  differential approximants to the series are shown.
The final estimate at $x=1$ is shown as a cross with error bar.
}
\label{fig_diffr}
\end{figure}

\begin{figure} 
\caption{Plot of one-hole dispersion
$\Delta/J$ (where $\Delta \equiv E^{\rm 1h}({\bf k})-E^{\rm 1h}(0, 0)$)
 along the path $\Gamma M X \Gamma$  
in the Brillouin zone (see Fig. \protect\ref{fig_lattice}(b)) 
for the system with coupling ratios
$t/J=$1/10, 1/5, 1/3, 1/2, 2/3 and 1.}
\label{fig_Ek1h}
\end{figure}

\begin{figure} 
\caption{Similar to Fig. \protect\ref{fig_Ek1h}, but with the hole dispersion 
curve plotted along the path
$\Gamma S M' \Gamma$ in the Brillouin zone.}
\label{fig_Ek1h2}
\end{figure}

\begin{figure} 
\caption{
The normalized bandwidth $W/t=[E^{1h}(0,0)-E^{1h}(\pi/2,\pi/2)]/t$
as a function of $J/t$. The solid square points with error bars connected
by a solid line are
our estimates from $x$-series, while the dashed line are our results
in Eq.(\protect\ref{W_2p}) obtained from the two-variable expansions.
The crosses are the estimates of
Martinez and Horsch\protect\cite{mar91}, the solid circles are the estimates
of Liu and Manousakis\protect\cite{liu92}.
}
\label{fig_W}
\end{figure}

\begin{figure} 
\caption{(a) The open (full) points represent the up (down) sublattices, respectively,
while the crosses represent the centre points of bonds connecting two adjacent sites.
The primitive translation vectors, $(e_x, e_y)$ and $(e_{x'}, e_{y'})$, are shown
by the arrows. (b) The reciprocal space of these lattices. The Brillouin zones for
$(e_x, e_y)$  and $(e_{x'}, e_{y'})$ are 
$-\protect\sqrt{2}\pi/a< k_x,k_y\leq \protect\sqrt{2} \pi/a$ 
and $-\pi/(\protect\sqrt{2}a)< k_{x'},k_{y'}\leq \pi/(\protect\sqrt{2}a)$, respectively.
}
\label{fig_lattice2h}
\end{figure}

\begin{figure} 
\caption{
Plot of 2-hole  dispersion
$E^{\rm 2h} ({\bf k})/J$
 along high-symmetry
cuts through the Brillouin zone for the system with coupling ratios
$J/t=10, 5, 3, 2, 3/2$.
}
\label{fig_Ek2h}
\end{figure}

\begin{figure} 
\caption{
Plot of the estimated binding energy
$E_{\rm b}/J$ for $s$-, $p$- and $d$-wave pair states (marked by
s, p and d, respectively) as  functions of $t/J$.
The crosses  with error bars connected by dashed lines 
are our estimates from $x$-series, 
while the solid lines in the small $t/J$ region 
are our results in Eq.(\protect\ref{eq_binding}) 
obtained from the two-variable expansions. The 
three solid circles with error bars are
 $8\times 8$ Green's function Monte Carlo results taken
from Ref. \protect\cite{bon}, 
the five solid triangles are the results of 32-site finite lattice
calculations\protect\cite{shih98,che98},
while the two open circles  are the 
extrapolation to bulk limit from finite size calculations\protect\cite{che98}.
}
\label{fig_binding}
\end{figure}

\begin{figure} 
\caption{Plot of the one-hole dispersion
$\Delta/J_z$ (where $\Delta \equiv E^{\rm 1h}({\bf k})-E^{\rm 1h}(0, 0)$)
along the path $\Gamma M X \Gamma$  
in the Brillouin zone (see Fig. \protect\ref{fig_lattice}(b)) 
 for the $t\hbox{-}J_z$ model with coupling ratios
$t/J_z$=5/3, 5/4, 1 and 0.8.}
\label{fig_Ek1h_tJz}
\end{figure}

\begin{figure} 
\caption{Plot of the one-hole energy at ${\bf k}=0$ for the $t-J_z$ model
as a function of 
$(J_z/t)^{2/3}/[1+(J_z/t)^{2/3}]$. The solid curves in the large
$(J_z/t)^{2/3}/[1+(J_z/t)^{2/3}]$ region are the integrated
differential approximants to the $t/J_z$ series, 
the points with error bars are estimates from $x$-series, while
the solid curve in the small $(J_z/t)^{2/3}/[1+(J_z/t)^{2/3}]$ region
is the least square fit $E^{1h}/t=-3.60+2.9(J_z/t)^{2/3} + J_z/t$ 
to the results in the interval $0.2 \leq J_z/t \leq 1$.
}
\label{fig_E1hmin}
\end{figure}

\begin{figure} 
\caption{
Plot of the 2-hole dispersion
$E^{\rm 2h} ({\bf k})/J_z$
 along high-symmetry
cuts through the Brillouin zone for the $t\hbox{-}J_z$ model with coupling ratios
$t/J_z=$0.2, 0.4, 0.6, 0.8 and 1.
}
\label{fig_Ek2s_tJz}
\end{figure}

\begin{figure} 
\caption{
Plot of the binding energy
$E_{\rm b}/J$ as a function of $t/J_z$ for the $t\hbox{-}J_z$ model.
The solid lines are the different integrated  differential approximants to 
the expansion in $t/J_z$, while the points with error bars connected by dashed lines
 are the estimates from the expansion in $x$.
The solid points are the results of  a finite size
calculations\protect\cite{rie93}.
}
\label{fig_binding_tJz}
\end{figure}


\setdec 0.00000000000
\begin{table}
\squeezetable
\caption{Series coefficients for the one-hole  energy 
 $E^{1h}/J_z =\sum_{k} c_{k} x^{k}$ for $y=0.5$  at ${\bf k}
=(\pi/2,\pi/2)$ for the $t\hbox{-}J$ model.
 Nonzero coefficients $c_{i}$
up to order $k=11$ for $r=0,1,2,4,8$ are listed.}\label{tab_1hmin}
\begin{tabular}{rrrrrr}
\multicolumn{1}{c}{$k$} &\multicolumn{1}{c}{$r=0$}
&\multicolumn{1}{c}{$r=1$} &\multicolumn{1}{c}{$r=2$}
&\multicolumn{1}{c}{$r=4$} &\multicolumn{1}{c}{$r=8$} \\
\hline
  0 &\dec  2.000000000   &\dec  2.500000000   &\dec  3.000000000   &\dec  4.000000000   &\dec  6.000000000   \\
  1 &\dec  0.000000000   &\dec $-$5.000000000$\times 10^{-1}$ &\dec $-$1.000000000   &\dec $-$2.000000000   &\dec $-$4.000000000   \\
  2 &\dec $-$5.333333333$\times 10^{-1}$ &\dec $-$2.666666667$\times 10^{-1}$ &\dec $-$1.758241758$\times 10^{-1}$ &\dec $-$1.038961039$\times 10^{-1}$ &\dec $-$5.689900427$\times 10^{-2}$ \\
  3 &\dec $-$2.666666667$\times 10^{-1}$ &\dec $-$2.251851852$\times 10^{-1}$ &\dec $-$1.647144065$\times 10^{-1}$ &\dec $-$1.027716872$\times 10^{-1}$ &\dec $-$5.729559761$\times 10^{-2}$ \\
  4 &\dec  2.062354497$\times 10^{-1}$ &\dec $-$8.432815051$\times 10^{-2}$ &\dec $-$1.116301281$\times 10^{-1}$ &\dec $-$8.974496745$\times 10^{-2}$ &\dec $-$5.525294748$\times 10^{-2}$ \\
  5 &\dec  2.121710758$\times 10^{-1}$ &\dec  1.942107294$\times 10^{-2}$ &\dec $-$5.439972676$\times 10^{-2}$ &\dec $-$7.119994116$\times 10^{-2}$ &\dec $-$5.149488552$\times 10^{-2}$ \\
  6 &\dec $-$2.058364193$\times 10^{-2}$ &\dec  4.770714647$\times 10^{-2}$ &\dec $-$1.307093131$\times 10^{-2}$ &\dec $-$5.177543994$\times 10^{-2}$ &\dec $-$4.663864987$\times 10^{-2}$ \\
  7 &\dec $-$2.275638025$\times 10^{-1}$ &\dec  2.527038112$\times 10^{-2}$ &\dec  7.461856693$\times 10^{-3}$ &\dec $-$3.440622079$\times 10^{-2}$ &\dec $-$4.119274206$\times 10^{-2}$ \\
  8 &\dec $-$9.889853918$\times 10^{-2}$ &\dec $-$7.079155979$\times 10^{-3}$ &\dec  1.149881952$\times 10^{-2}$ &\dec $-$2.058064339$\times 10^{-2}$ &\dec $-$3.556056761$\times 10^{-2}$ \\
  9 &\dec  1.839605467$\times 10^{-1}$ &\dec $-$2.298423843$\times 10^{-2}$ &\dec  6.622730013$\times 10^{-3}$ &\dec $-$1.069884345$\times 10^{-2}$ &\dec $-$3.004823602$\times 10^{-2}$ \\
 10 &\dec  2.267402233$\times 10^{-1}$ &\dec $-$1.850514883$\times 10^{-2}$ &\dec $-$3.889599620$\times 10^{-4}$ &\dec $-$4.444287027$\times 10^{-3}$ &\dec $-$2.487508963$\times 10^{-2}$ \\
 11 &\dec $-$9.850744312$\times 10^{-2}$ &\dec $-$4.463638360$\times 10^{-3}$ &\dec $-$5.435829342$\times 10^{-3}$ &\dec $-$1.112182682$\times 10^{-3}$ &\dec $-$2.018574494$\times 10^{-2}$ \\
\end{tabular}
\end{table}

\setdec 0.00000000000
\begin{table}
\squeezetable
\caption{Series coefficients for the one-hole and two-hole  energy 
 $E (k_x, k_y)/J_z =\sum_{k,n,m} a_{k,n,m}$ $x^{k}
 [\cos (n k_x A) \cos (m k_y A) + \cos (m k_x A) \cos (n k_y A) ]/2$ of the $t\hbox{-}J$ model,
where $A=a$ for the one-hole case and $A=a/\protect\sqrt{2}$ for the two-hole case.
 Nonzero coefficients $a_{k,n,m}$
up to order $k=11$ for $y=0.5$ and $r=0$ are listed.}\label{tab_tjK}
\begin{tabular}{rr|rr|rr|rr}
\multicolumn{1}{c}{(k,n,m)} &\multicolumn{1}{c|}{$a_{k,n,m}$}
&\multicolumn{1}{c}{(k,n,m)} &\multicolumn{1}{c|}{$a_{k,n,m}$}
&\multicolumn{1}{c}{(k,n,m)} &\multicolumn{1}{c|}{$a_{k,n,m}$}
&\multicolumn{1}{c}{(k,n,m)} &\multicolumn{1}{c}{$a_{k,n,m}$} \\
\hline
\multicolumn{8}{c}{one hole energy $E^{1h}(k_x, k_y)/J_z$} \\
 ( 0, 0, 0) &\dec   2.000000000   &( 8, 2, 0) &\dec  -8.693161815$\times 10^{-2}$ &( 5, 1, 1) &\dec  -3.650599647$\times 10^{-1}$ &(11, 5, 1) &\dec  -2.756493796$\times 10^{-2}$ \\
 ( 2, 0, 0) &\dec  -5.333333333$\times 10^{-1}$ &( 9, 2, 0) &\dec   2.176985203$\times 10^{-1}$ &( 6, 1, 1) &\dec   2.714260859$\times 10^{-1}$ &( 6, 2, 2) &\dec   7.130158730$\times 10^{-2}$ \\
 ( 4, 0, 0) &\dec   1.529021164$\times 10^{-1}$ &(10, 2, 0) &\dec  -1.017391598$\times 10^{-2}$ &( 7, 1, 1) &\dec   2.890627842$\times 10^{-1}$ &( 7, 2, 2) &\dec  -6.731669606$\times 10^{-2}$ \\
 ( 5, 0, 0) &\dec   9.481481481$\times 10^{-2}$ &(11, 2, 0) &\dec  -3.628184110$\times 10^{-1}$ &( 8, 1, 1) &\dec  -2.000742064$\times 10^{-1}$ &( 8, 2, 2) &\dec  -1.075223788$\times 10^{-1}$ \\
 ( 6, 0, 0) &\dec  -1.772367448$\times 10^{-2}$ &( 6, 4, 0) &\dec   1.188359788$\times 10^{-2}$ &( 9, 1, 1) &\dec  -1.603987396$\times 10^{-1}$ &( 9, 2, 2) &\dec   1.539341160$\times 10^{-1}$ \\
 ( 7, 0, 0) &\dec  -1.684678469$\times 10^{-1}$ &( 7, 4, 0) &\dec  -4.980599647$\times 10^{-3}$ &(10, 1, 1) &\dec  -1.153947213$\times 10^{-1}$ &(10, 2, 2) &\dec   9.103499935$\times 10^{-2}$ \\
 ( 8, 0, 0) &\dec  -6.427987242$\times 10^{-2}$ &( 8, 4, 0) &\dec  -1.402790612$\times 10^{-2}$ &(11, 1, 1) &\dec   1.409738327$\times 10^{-1}$ &(11, 2, 2) &\dec  -2.080643863$\times 10^{-1}$ \\
 ( 9, 0, 0) &\dec   2.538548842$\times 10^{-1}$ &( 9, 4, 0) &\dec   2.637981607$\times 10^{-2}$ &( 6, 3, 1) &\dec   9.506878307$\times 10^{-2}$ &( 9, 4, 2) &\dec   3.146104767$\times 10^{-2}$ \\
 (10, 0, 0) &\dec   1.029816377$\times 10^{-1}$ &(10, 4, 0) &\dec  -1.865761552$\times 10^{-2}$ &( 7, 3, 1) &\dec  -7.727789536$\times 10^{-2}$ &(10, 4, 2) &\dec  -4.054217598$\times 10^{-2}$ \\
 (11, 0, 0) &\dec  -2.786820323$\times 10^{-1}$ &(11, 4, 0) &\dec  -4.652610741$\times 10^{-2}$ &( 8, 3, 1) &\dec  -1.457528700$\times 10^{-1}$ &(11, 4, 2) &\dec  -7.000652266$\times 10^{-2}$ \\
 ( 3, 2, 0) &\dec   2.666666667$\times 10^{-1}$ &( 9, 6, 0) &\dec   1.048701589$\times 10^{-3}$ &( 9, 3, 1) &\dec   1.806575878$\times 10^{-1}$ &( 9, 3, 3) &\dec   2.097403178$\times 10^{-2}$ \\
 ( 4, 2, 0) &\dec  -5.333333333$\times 10^{-2}$ &(10, 6, 0) &\dec  -6.651097800$\times 10^{-4}$ &(10, 3, 1) &\dec   8.928712117$\times 10^{-2}$ &(10, 3, 3) &\dec  -2.874385753$\times 10^{-2}$ \\
 ( 5, 2, 0) &\dec  -1.173562610$\times 10^{-1}$ &(11, 6, 0) &\dec  -1.940149264$\times 10^{-3}$ &(11, 3, 1) &\dec  -2.245856051$\times 10^{-1}$ &(11, 3, 3) &\dec  -4.632188322$\times 10^{-2}$ \\
 ( 6, 2, 0) &\dec   8.604515263$\times 10^{-2}$ &( 3, 1, 1) &\dec   5.333333333$\times 10^{-1}$ &( 9, 5, 1) &\dec   1.258441907$\times 10^{-2}$  \\
 ( 7, 2, 0) &\dec  -1.320134011$\times 10^{-2}$ &( 4, 1, 1) &\dec  -2.962962963$\times 10^{-1}$ &(10, 5, 1) &\dec  -1.312853800$\times 10^{-2}$  \\
\hline
\multicolumn{8}{c}{two-hole energy $E^{2h}(k_x, k_y)/J_z$} \\
 ( 0, 0, 0) &\dec   3.500000000  &( 6, 3, 3) &\dec   1.457965902$\times 10^{-3}$ &( 6, 3, 2) &\dec   5.831863610$\times 10^{-3}$ &( 9, 4, 2) &\dec $-$7.128118497$\times 10^{-4}$  \\
 ( 2, 0, 0) &\dec $-$7.333333333$\times 10^{-1}$ &( 8, 3, 3) &\dec $-$8.399019298$\times 10^{-3}$ &( 7, 3, 2) &\dec $-$4.705118313$\times 10^{-3}$ &(10, 4, 2) &\dec $-$6.158730192$\times 10^{-2}$  \\
 ( 4, 0, 0) &\dec   1.945310847$\times 10^{-1}$ &( 9, 3, 3) &\dec   2.106020257$\times 10^{-2}$ &( 8, 3, 2) &\dec $-$5.937350737$\times 10^{-2}$ &(11, 4, 2) &\dec   6.702194733$\times 10^{-2}$  \\
 ( 5, 0, 0) &\dec   1.570370370$\times 10^{-1}$ &(10, 3, 3) &\dec   5.953632826$\times 10^{-3}$ &( 9, 3, 2) &\dec   3.295758313$\times 10^{-2}$ &(10, 5, 3) &\dec   2.643901061$\times 10^{-4}$  \\
 ( 6, 0, 0) &\dec $-$1.814429222$\times 10^{-1}$ &(11, 3, 3) &\dec   4.246104998$\times 10^{-3}$ &(10, 3, 2) &\dec   1.653086795$\times 10^{-2}$ &(11, 5, 3) &\dec $-$8.754870105$\times 10^{-5}$  \\
 ( 7, 0, 0) &\dec   3.336368382$\times 10^{-2}$ &( 8, 4, 4) &\dec   1.843920010$\times 10^{-4}$ &(11, 3, 2) &\dec $-$1.386696140$\times 10^{-2}$ &( 6, 3, 0) &\dec   5.831863610$\times 10^{-3}$  \\
 ( 8, 0, 0) &\dec $-$1.977570791$\times 10^{-1}$ &(10, 4, 4) &\dec $-$3.073373856$\times 10^{-3}$ &( 8, 4, 3) &\dec   7.375680039$\times 10^{-4}$ &( 7, 3, 0) &\dec $-$1.162496326$\times 10^{-3}$  \\
 ( 9, 0, 0) &\dec $-$1.214962788$\times 10^{-1}$ &(11, 4, 4) &\dec   4.127194476$\times 10^{-4}$ &( 9, 4, 3) &\dec $-$5.394036567$\times 10^{-4}$ &( 8, 3, 0) &\dec $-$1.040888381$\times 10^{-1}$  \\
 (10, 0, 0) &\dec   1.897994391$\times 10^{-1}$ &(10, 5, 5) &\dec   2.643901061$\times 10^{-5}$ &(10, 4, 3) &\dec $-$2.344210680$\times 10^{-2}$ &( 9, 3, 0) &\dec   8.179624689$\times 10^{-2}$  \\
 (11, 0, 0) &\dec   7.710636344$\times 10^{-1}$ &( 2, 1, 0) &\dec   6.666666667$\times 10^{-1}$ &(11, 4, 3) &\dec   3.300423401$\times 10^{-2}$ &(10, 3, 0) &\dec   1.171660236$\times 10^{-1}$  \\
 ( 2, 1, 1) &\dec   3.333333333$\times 10^{-1}$ &( 3, 1, 0) &\dec   3.333333333$\times 10^{-1}$ &(10, 5, 4) &\dec   1.057560424$\times 10^{-4}$ &(11, 3, 0) &\dec   1.531781787$\times 10^{-2}$  \\
 ( 3, 1, 1) &\dec $-$3.333333333$\times 10^{-1}$ &( 4, 1, 0) &\dec $-$2.673809524$\times 10^{-1}$ &(11, 5, 4) &\dec $-$7.959003190$\times 10^{-5}$ &( 8, 4, 1) &\dec   2.212704012$\times 10^{-3}$  \\
 ( 4, 1, 1) &\dec $-$1.192857143$\times 10^{-1}$ &( 5, 1, 0) &\dec $-$4.024977954$\times 10^{-1}$ &( 4, 2, 0) &\dec   2.962962963$\times 10^{-2}$ &( 9, 4, 1) &\dec $-$8.862200427$\times 10^{-4}$  \\
 ( 5, 1, 1) &\dec   2.904905203$\times 10^{-1}$ &( 6, 1, 0) &\dec $-$1.302727805$\times 10^{-1}$ &( 5, 2, 0) &\dec $-$2.629629630$\times 10^{-2}$ &(10, 4, 1) &\dec $-$9.978608234$\times 10^{-2}$  \\
 ( 6, 1, 1) &\dec   5.329329186$\times 10^{-1}$ &( 7, 1, 0) &\dec   3.499536691$\times 10^{-1}$ &( 6, 2, 0) &\dec $-$2.375711537$\times 10^{-1}$ &(11, 4, 1) &\dec   1.153622881$\times 10^{-1}$  \\
 ( 7, 1, 1) &\dec $-$6.271803816$\times 10^{-1}$ &( 8, 1, 0) &\dec   4.079418136$\times 10^{-1}$ &( 7, 2, 0) &\dec   2.330090787$\times 10^{-1}$ &(10, 5, 2) &\dec   4.230241698$\times 10^{-4}$  \\
 ( 8, 1, 1) &\dec $-$7.522337850$\times 10^{-1}$ &( 9, 1, 0) &\dec   5.301735104$\times 10^{-2}$ &( 8, 2, 0) &\dec   3.039557383$\times 10^{-1}$ &(11, 5, 2) &\dec $-$1.750974021$\times 10^{-4}$  \\
 ( 9, 1, 1) &\dec   5.347837206$\times 10^{-1}$ &(10, 1, 0) &\dec $-$1.026434348  &( 9, 2, 0) &\dec $-$3.529882617$\times 10^{-1}$ &( 8, 4, 0) &\dec   1.106352006$\times 10^{-3}$  \\
 (10, 1, 1) &\dec   1.146091910  &(11, 1, 0) &\dec $-$7.412817539$\times 10^{-1}$ &(10, 2, 0) &\dec $-$6.813379022$\times 10^{-1}$ &( 9, 4, 0) &\dec $-$7.128118497$\times 10^{-4}$  \\
 (11, 1, 1) &\dec $-$4.863394606$\times 10^{-1}$ &( 4, 2, 1) &\dec   5.925925926$\times 10^{-2}$ &(11, 2, 0) &\dec   9.790814131$\times 10^{-1}$ &(10, 4, 0) &\dec $-$5.538838348$\times 10^{-2}$  \\
 ( 4, 2, 2) &\dec   1.481481481$\times 10^{-2}$ &( 5, 2, 1) &\dec $-$2.629629630$\times 10^{-2}$ &( 6, 3, 1) &\dec   8.747795414$\times 10^{-3}$ &(11, 4, 0) &\dec   8.043954590$\times 10^{-2}$  \\
 ( 6, 2, 2) &\dec $-$5.533170820$\times 10^{-2}$ &( 6, 2, 1) &\dec $-$8.610412317$\times 10^{-2}$ &( 7, 3, 1) &\dec $-$5.867614638$\times 10^{-3}$ &(10, 5, 1) &\dec   5.287802122$\times 10^{-4}$  \\
 ( 7, 2, 2) &\dec   1.147526740$\times 10^{-2}$ &( 7, 2, 1) &\dec   5.769980192$\times 10^{-2}$ &( 8, 3, 1) &\dec $-$1.459267389$\times 10^{-1}$ &(11, 5, 1) &\dec $-$2.626461032$\times 10^{-4}$  \\
 ( 8, 2, 2) &\dec   3.484612337$\times 10^{-2}$ &( 8, 2, 1) &\dec   2.118965896$\times 10^{-1}$ &( 9, 3, 1) &\dec   1.611775400$\times 10^{-1}$ &(10, 5, 0) &\dec   3.172681273$\times 10^{-4}$  \\
 ( 9, 2, 2) &\dec $-$8.489781072$\times 10^{-2}$ &( 9, 2, 1) &\dec $-$9.184491129$\times 10^{-2}$ &(10, 3, 1) &\dec   1.623787831$\times 10^{-1}$ &(11, 5, 0) &\dec $-$9.550737020$\times 10^{-5}$  \\
 (10, 2, 2) &\dec $-$8.365174384$\times 10^{-2}$ &(10, 2, 1) &\dec $-$3.638022235$\times 10^{-1}$ &(11, 3, 1) &\dec $-$3.687197934$\times 10^{-1}$ \\
 (11, 2, 2) &\dec   3.945274143$\times 10^{-1}$ &(11, 2, 1) &\dec $-$5.611998341$\times 10^{-3}$ &( 8, 4, 2) &\dec   1.475136008$\times 10^{-3}$  \\
\end{tabular}
\end{table}

\setdec 0.00000000000
\begin{table}
\squeezetable
\caption{Series coefficients for the one-hole and two-hole energy 
 $E (k_x, k_y)/J_z =\sum_{k,n,m} $ $a_{k,n,m}$ $(t/J_z)^{k}
 [\cos (n k_x A) \cos (m k_y A) + \cos (m k_x A) \cos (n k_y A) ]/2$ for the
Hamiltonian in Eq. (\protect\ref{H_tJz}) of the $t\hbox{-}J_z$ model,
where $A=a$ for the one-hole case and $A=a/\protect\sqrt{2}$ for the two-hole case.
 Nonzero coefficients $a_{k,n,m}$
up to order $k=20$ for one-hole case or $k=18$ for two-hole case  are listed.}\label{tabtJz}
\begin{tabular}{rr|rr|rr|rr}
\multicolumn{1}{c}{(k,n,m)} &\multicolumn{1}{c|}{$a_{k,n,m}$}
&\multicolumn{1}{c}{(k,n,m)} &\multicolumn{1}{c|}{$a_{k,n,m}$}
&\multicolumn{1}{c}{(k,n,m)} &\multicolumn{1}{c|}{$a_{k,n,m}$}
&\multicolumn{1}{c}{(k,n,m)} &\multicolumn{1}{c}{$a_{k,n,m}$} \\
\hline
\multicolumn{8}{c}{one hole energy $E^{1h}(k_x, k_y)/J_z$} \\
 ( 0, 0, 0) &\dec   2.000000000 &(20, 0, 0) &\dec   3.269830958$\times 10^{4}$ &( 8, 1, 1) &\dec   1.107677837 &(20, 3, 1) &\dec   1.363566623  \\
 ( 2, 0, 0) &\dec $-$2.666666667 &(10, 2, 0) &\dec $-$9.187011686$\times 10^{-3}$ &(10, 1, 1) &\dec $-$5.258263841 &(12, 2, 2) &\dec   2.213826787$\times 10^{-2}$  \\
 ( 4, 0, 0) &\dec   2.607407407 &(12, 2, 0) &\dec   1.237459977$\times 10^{-1}$ &(12, 1, 1) &\dec   2.361854957$\times 10^{1}$ &(14, 2, 2) &\dec $-$2.801396443$\times 10^{-1}$  \\
 ( 6, 0, 0) &\dec $-$5.667818930 &(14, 2, 0) &\dec $-$1.042288265 &(14, 1, 1) &\dec $-$1.041949744$\times 10^{2}$ &(16, 2, 2) &\dec   2.293697931  \\
 ( 8, 0, 0) &\dec   1.579276071$\times 10^{1}$ &(16, 2, 0) &\dec   7.139692772 &(16, 1, 1) &\dec   4.570753372$\times 10^{2}$ &(18, 2, 2) &\dec $-$1.550982309$\times 10^{1}$  \\
 (10, 0, 0) &\dec $-$4.977461910$\times 10^{1}$ &(18, 2, 0) &\dec $-$4.367374242$\times 10^{1}$ &(18, 1, 1) &\dec $-$2.003192131$\times 10^{3}$ &(20, 2, 2) &\dec   9.435217647$\times 10^{1}$  \\
 (12, 0, 0) &\dec   1.689229038$\times 10^{2}$ &(20, 2, 0) &\dec   2.492266705$\times 10^{2}$ &(20, 1, 1) &\dec   8.787977137$\times 10^{3}$ &(20, 4, 2) &\dec   4.053738087$\times 10^{-5}$  \\
 (14, 0, 0) &\dec $-$6.023423934$\times 10^{2}$ &(18, 4, 0) &\dec $-$1.275445883$\times 10^{-6}$ &(14, 3, 1) &\dec $-$1.112602736$\times 10^{-4}$ &(18, 3, 3) &\dec $-$4.160122546$\times 10^{-3}$  \\
 (16, 0, 0) &\dec   2.225264150$\times 10^{3}$ &(20, 4, 0) &\dec   1.432818794$\times 10^{-4}$ &(16, 3, 1) &\dec   6.891013162$\times 10^{-3}$ &(20, 3, 3) &\dec   8.089745849$\times 10^{-2}$  \\
 (18, 0, 0) &\dec $-$8.447437043$\times 10^{3}$ &( 6, 1, 1) &\dec $-$1.896296296$\times 10^{-1}$ &(18, 3, 1) &\dec $-$1.208165780$\times 10^{-1}$  \\
\hline
\multicolumn{8}{c}{two-hole energy $E^{2h}(k_x, k_y)/J_z$} \\
 ( 0, 0, 0) &\dec   3.500000000  &( 2, 1, 0) &\dec   2.666666667  &(10, 3, 1) &\dec $-$3.833568660  &(10, 4, 0) &\dec $-$4.187749256$\times 10^{-1}$  \\
 ( 2, 0, 0) &\dec $-$4.000000000  &( 4, 1, 0) &\dec $-$2.607407407  &(12, 3, 1) &\dec   3.749428458$\times 10^{1}$ &(12, 4, 0) &\dec $-$4.625682644   \\
 ( 4, 0, 0) &\dec   2.311111111  &( 6, 1, 0) &\dec $-$9.124044680$\times 10^{-1}$ &(14, 3, 1) &\dec $-$1.763568623$\times 10^{2}$ &(14, 4, 0) &\dec   3.879009243$\times 10^{1}$  \\
 ( 6, 0, 0) &\dec $-$3.578130511$\times 10^{-1}$ &( 8, 1, 0) &\dec   2.709462265$\times 10^{1}$ &(16, 3, 1) &\dec   5.362296757$\times 10^{2}$ &(16, 4, 0) &\dec $-$1.755530196$\times 10^{2}$  \\
 ( 8, 0, 0) &\dec $-$1.447791909$\times 10^{1}$ &(10, 1, 0) &\dec $-$1.454354060$\times 10^{2}$ &(18, 3, 1) &\dec $-$4.338040480$\times 10^{2}$ &(18, 4, 0) &\dec   5.141129944$\times 10^{2}$  \\
 (10, 0, 0) &\dec   8.449051521$\times 10^{1}$ &(12, 1, 0) &\dec   4.973154521$\times 10^{2}$ &( 8, 4, 2) &\dec   3.776348180$\times 10^{-1}$ &(10, 5, 1) &\dec   5.414709373$\times 10^{-1}$  \\
 (12, 0, 0) &\dec $-$3.205718341$\times 10^{2}$ &(14, 1, 0) &\dec $-$8.044807264$\times 10^{2}$ &(10, 4, 2) &\dec $-$7.694903208$\times 10^{-1}$ &(12, 5, 1) &\dec $-$6.844512414$\times 10^{-1}$  \\
 (14, 0, 0) &\dec   7.683946158$\times 10^{2}$ &(16, 1, 0) &\dec $-$3.592648436$\times 10^{3}$ &(12, 4, 2) &\dec $-$3.834200865  &(14, 5, 1) &\dec $-$1.368955321$\times 10^{1}$  \\
 (16, 0, 0) &\dec   1.864995221$\times 10^{2}$ &(18, 1, 0) &\dec   4.011882196$\times 10^{4}$ &(14, 4, 2) &\dec   4.301073460$\times 10^{1}$ &(16, 5, 1) &\dec   1.169642314$\times 10^{2}$  \\
 (18, 0, 0) &\dec $-$1.447204343$\times 10^{4}$ &( 4, 2, 1) &\dec   9.481481481$\times 10^{-1}$ &(16, 4, 2) &\dec $-$2.363456039$\times 10^{2}$ &(18, 5, 1) &\dec $-$5.467419040$\times 10^{2}$  \\
 ( 2, 1, 1) &\dec   1.333333333  &( 6, 2, 1) &\dec $-$2.072004703  &(18, 4, 2) &\dec   9.094821992$\times 10^{2}$ &(12, 6, 2) &\dec   5.024245295$\times 10^{-1}$  \\
 ( 4, 1, 1) &\dec $-$1.362962963  &( 8, 2, 1) &\dec   1.071796284  &(10, 5, 3) &\dec   2.707354687$\times 10^{-1}$ &(14, 6, 2) &\dec $-$5.607740036$\times 10^{-1}$  \\
 ( 6, 1, 1) &\dec   3.117601411  &(10, 2, 1) &\dec   1.934131152$\times 10^{1}$ &(12, 5, 3) &\dec $-$6.099856791$\times 10^{-1}$ &(16, 6, 2) &\dec $-$1.688448287$\times 10^{1}$  \\
 ( 8, 1, 1) &\dec $-$1.157205899$\times 10^{1}$ &(12, 2, 1) &\dec $-$1.610227509$\times 10^{2}$ &(14, 5, 3) &\dec $-$3.591949866  &(18, 6, 2) &\dec   1.530376091$\times 10^{2}$  \\
 (10, 1, 1) &\dec   3.724437583$\times 10^{1}$ &(14, 2, 1) &\dec   9.100666397$\times 10^{2}$ &(16, 5, 3) &\dec   4.680082984$\times 10^{1}$ &(14, 7, 3) &\dec   4.576926174$\times 10^{-1}$  \\
 (12, 1, 1) &\dec $-$4.436033620$\times 10^{1}$ &(16, 2, 1) &\dec $-$4.053430224$\times 10^{3}$ &(18, 5, 3) &\dec $-$2.944841481$\times 10^{2}$ &(16, 7, 3) &\dec $-$4.694339733$\times 10^{-1}$  \\
 (14, 1, 1) &\dec $-$5.334708723$\times 10^{2}$ &(18, 2, 1) &\dec   1.279544679$\times 10^{4}$ &(12, 6, 4) &\dec   2.009698118$\times 10^{-1}$ &(18, 7, 3) &\dec $-$1.886986003$\times 10^{1}$  \\
 (16, 1, 1) &\dec   5.421019234$\times 10^{3}$ &( 6, 3, 2) &\dec   3.732392710$\times 10^{-1}$ &(14, 6, 4) &\dec $-$4.925030503$\times 10^{-1}$ &(16, 8, 4) &\dec   4.117367040$\times 10^{-1}$  \\
 (18, 1, 1) &\dec $-$3.093983511$\times 10^{4}$ &( 8, 3, 2) &\dec $-$7.739781828$\times 10^{-1}$ &(16, 6, 4) &\dec $-$3.212472559  &(18, 8, 4) &\dec $-$4.052076547$\times 10^{-1}$  \\
 ( 4, 2, 2) &\dec   2.370370370$\times 10^{-1}$ &(10, 3, 2) &\dec $-$1.053007756  &(18, 6, 4) &\dec   4.903204193$\times 10^{1}$ &(18, 9, 5) &\dec   3.670192948$\times 10^{-1}$  \\
 ( 6, 2, 2) &\dec $-$5.237389771$\times 10^{-1}$ &(12, 3, 2) &\dec   1.967483048$\times 10^{1}$ &(14, 7, 5) &\dec   1.525642058$\times 10^{-1}$ &(10, 5, 0) &\dec   3.248825624$\times 10^{-1}$  \\
 ( 8, 2, 2) &\dec   1.010441376  &(14, 3, 2) &\dec $-$1.305817944$\times 10^{2}$ &(16, 7, 5) &\dec $-$4.034853659$\times 10^{-1}$ &(12, 5, 0) &\dec $-$6.899424873$\times 10^{-1}$  \\
 (10, 2, 2) &\dec $-$1.494008026  &(16, 3, 2) &\dec   6.744405152$\times 10^{2}$ &(18, 7, 5) &\dec $-$2.770916581  &(14, 5, 0) &\dec $-$6.463536863   \\
 (12, 2, 2) &\dec   1.846909265$\times 10^{-1}$ &(18, 3, 2) &\dec $-$2.972249754$\times 10^{3}$ &(16, 8, 6) &\dec   1.176390583$\times 10^{-1}$ &(16, 5, 0) &\dec   6.567477615$\times 10^{1}$  \\
 (14, 2, 2) &\dec   1.042676508$\times 10^{1}$ &( 8, 4, 3) &\dec   1.888174090$\times 10^{-1}$ &(18, 8, 6) &\dec $-$3.342928209$\times 10^{-1}$ &(18, 5, 0) &\dec $-$3.388841937$\times 10^{2}$  \\
 (16, 2, 2) &\dec $-$1.111625374$\times 10^{2}$ &(10, 4, 3) &\dec $-$2.886375899$\times 10^{-1}$ &(18, 9, 7) &\dec   9.175482371$\times 10^{-2}$ &(12, 6, 1) &\dec   6.698993726$\times 10^{-1}$  \\
 (18, 2, 2) &\dec   1.249271047$\times 10^{3}$ &(12, 4, 3) &\dec $-$1.577708121  &( 6, 3, 0) &\dec   3.732392710$\times 10^{-1}$ &(14, 6, 1) &\dec $-$1.079421253   \\
 ( 6, 3, 3) &\dec   9.330981775$\times 10^{-2}$ &(14, 4, 3) &\dec   1.855760450$\times 10^{1}$ &( 8, 3, 0) &\dec $-$9.806190570$\times 10^{-1}$ &(16, 6, 1) &\dec $-$2.116718725$\times 10^{1}$  \\
 ( 8, 3, 3) &\dec $-$2.869718840$\times 10^{-1}$ &(16, 4, 3) &\dec $-$1.206715261$\times 10^{2}$ &(10, 3, 0) &\dec $-$7.606578761$\times 10^{-1}$ &(18, 6, 1) &\dec   2.016971553$\times 10^{2}$  \\
 (10, 3, 3) &\dec   7.302649849$\times 10^{-1}$ &(18, 4, 3) &\dec   6.228213110$\times 10^{2}$ &(12, 3, 0) &\dec   1.900360917$\times 10^{1}$ &(14, 7, 2) &\dec   6.538465963$\times 10^{-1}$  \\
 (12, 3, 3) &\dec   6.474454323$\times 10^{-2}$ &(10, 5, 4) &\dec   1.082941875$\times 10^{-1}$ &(14, 3, 0) &\dec $-$1.096944607$\times 10^{2}$ &(16, 7, 2) &\dec $-$6.381513372$\times 10^{-1}$  \\
 (14, 3, 3) &\dec $-$1.507976852$\times 10^{1}$ &(12, 5, 4) &\dec $-$7.182655253$\times 10^{-2}$ &(16, 3, 0) &\dec   4.525126033$\times 10^{2}$ &(18, 7, 2) &\dec $-$2.878600133$\times 10^{1}$  \\
 (16, 3, 3) &\dec   1.112739551$\times 10^{2}$ &(14, 5, 4) &\dec $-$1.662709009  &(18, 3, 0) &\dec $-$1.604127694$\times 10^{3}$ &(16, 8, 3) &\dec   6.176050560$\times 10^{-1}$  \\
 (18, 3, 3) &\dec $-$5.501017323$\times 10^{2}$ &(16, 5, 4) &\dec   1.713732791$\times 10^{1}$ &( 8, 4, 1) &\dec   5.664522270$\times 10^{-1}$ &(18, 8, 3) &\dec $-$1.660697208$\times 10^{-1}$  \\
 ( 8, 4, 4) &\dec   4.720435225$\times 10^{-2}$ &(18, 5, 4) &\dec $-$1.134996640$\times 10^{2}$ &(10, 4, 1) &\dec $-$1.358637239  &(18, 9, 4) &\dec   5.709189031$\times 10^{-1}$  \\
 (10, 4, 4) &\dec $-$1.753576976$\times 10^{-1}$ &(12, 6, 5) &\dec   6.698993726$\times 10^{-2}$ &(12, 4, 1) &\dec $-$5.476288758  &(12, 6, 0) &\dec   3.349496863$\times 10^{-1}$  \\
 (12, 4, 4) &\dec   5.665451332$\times 10^{-1}$ &(14, 6, 5) &\dec   2.348849413$\times 10^{-2}$ &(14, 4, 1) &\dec   6.555818772$\times 10^{1}$ &(14, 6, 0) &\dec $-$2.248529752$\times 10^{-1}$  \\
 (14, 4, 4) &\dec   7.031489323$\times 10^{-1}$ &(16, 6, 5) &\dec $-$1.498021733  &(16, 4, 1) &\dec $-$3.561148091$\times 10^{2}$ &(16, 6, 0) &\dec $-$1.304144759$\times 10^{1}$  \\
 (16, 4, 4) &\dec $-$2.234394856$\times 10^{1}$ &(18, 6, 5) &\dec   1.514875202$\times 10^{1}$ &(18, 4, 1) &\dec   1.327749407$\times 10^{3}$ &(18, 6, 0) &\dec   1.069591051$\times 10^{2}$  \\
 (18, 4, 4) &\dec   1.696295810$\times 10^{2}$ &(14, 7, 6) &\dec   4.358977309$\times 10^{-2}$ &(10, 5, 2) &\dec   4.331767498$\times 10^{-1}$ &(14, 7, 1) &\dec   7.628210290$\times 10^{-1}$  \\
 (10, 5, 5) &\dec   2.707354687$\times 10^{-2}$ &(16, 7, 6) &\dec   6.397044909$\times 10^{-2}$ &(12, 5, 2) &\dec $-$8.287589771$\times 10^{-1}$ &(16, 7, 1) &\dec $-$1.648715187$\times 10^{-1}$  \\
 (12, 5, 5) &\dec $-$1.145505796$\times 10^{-1}$ &(18, 7, 6) &\dec $-$1.267992012  &(14, 5, 2) &\dec $-$8.117281873  &(18, 7, 1) &\dec $-$3.918669310$\times 10^{1}$  \\
 (14, 5, 5) &\dec   4.544498080$\times 10^{-1}$ &(16, 8, 7) &\dec   2.940976457$\times 10^{-2}$ &(16, 5, 2) &\dec   8.399862105$\times 10^{1}$ &(16, 8, 2) &\dec   8.234734079$\times 10^{-1}$  \\
 (16, 5, 5) &\dec   1.078497914  &(18, 8, 7) &\dec   7.770975088$\times 10^{-2}$ &(18, 5, 2) &\dec $-$4.609565870$\times 10^{2}$ &(18, 8, 2) &\dec   1.073653528$\times 10^{-1}$  \\
 (18, 5, 5) &\dec $-$2.724985485$\times 10^{1}$ &(18, 9, 8) &\dec   2.038996082$\times 10^{-2}$ &(12, 6, 3) &\dec   3.349496863$\times 10^{-1}$ &(18, 9, 3) &\dec   8.563783546$\times 10^{-1}$  \\
 (12, 6, 6) &\dec   1.674748432$\times 10^{-2}$ &( 4, 2, 0) &\dec   4.740740741$\times 10^{-1}$ &(14, 6, 3) &\dec $-$4.223690648$\times 10^{-1}$ &(14, 7, 0) &\dec   4.358977309$\times 10^{-1}$  \\
 (14, 6, 6) &\dec $-$7.829101098$\times 10^{-2}$ &( 6, 2, 0) &\dec $-$6.512874780$\times 10^{-1}$ &(16, 6, 3) &\dec $-$9.671323639  &(16, 7, 0) &\dec $-$4.703472705$\times 10^{-1}$  \\
 (16, 6, 6) &\dec   3.741008913$\times 10^{-1}$ &( 8, 2, 0) &\dec $-$1.556811959  &(18, 6, 3) &\dec   9.548744493$\times 10^{1}$ &(18, 7, 0) &\dec $-$1.979896930$\times 10^{1}$  \\
 (18, 6, 6) &\dec   1.267008305  &(10, 2, 0) &\dec   1.207544510$\times 10^{1}$ &(14, 7, 4) &\dec   2.615386385$\times 10^{-1}$ &(16, 8, 1) &\dec   1.029341760   \\
 (14, 7, 7) &\dec   1.089744327$\times 10^{-2}$ &(12, 2, 0) &\dec $-$3.196277031$\times 10^{1}$ &(16, 7, 4) &\dec $-$1.332433821$\times 10^{-1}$ &(18, 8, 1) &\dec $-$4.479185205$\times 10^{-1}$  \\
 (16, 7, 7) &\dec $-$5.528545914$\times 10^{-2}$ &(14, 2, 0) &\dec $-$5.271684834$\times 10^{1}$ &(18, 7, 4) &\dec $-$1.033273379$\times 10^{1}$ &(18, 9, 2) &\dec   1.141837806   \\
 (18, 7, 7) &\dec   3.138641071$\times 10^{-1}$ &(16, 2, 0) &\dec   1.018254901$\times 10^{3}$ &(16, 8, 5) &\dec   2.058683520$\times 10^{-1}$ &(16, 8, 0) &\dec   5.146708800$\times 10^{-1}$  \\
 (16, 8, 8) &\dec   7.352441142$\times 10^{-3}$ &(18, 2, 0) &\dec $-$5.459613629$\times 10^{3}$ &(18, 8, 5) &\dec   6.203359335$\times 10^{-2}$ &(18, 8, 0) &\dec   2.583076501$\times 10^{-1}$  \\
 (18, 8, 8) &\dec $-$4.001373177$\times 10^{-2}$ &( 6, 3, 1) &\dec   5.598589065$\times 10^{-1}$ &(18, 9, 6) &\dec   1.631196866$\times 10^{-1}$ &(18, 9, 1) &\dec   1.284567532   \\
 (18, 9, 9) &\dec   5.097490206$\times 10^{-3}$ &( 8, 3, 1) &\dec $-$9.918360628$\times 10^{-1}$ &( 8, 4, 0) &\dec   2.832261135$\times 10^{-1}$ &(18, 9, 0) &\dec   7.136486288$\times 10^{-1}$  \\
\end{tabular}
\end{table}


\begin{references}
\bibitem[*]{byline1}e-mail address: C.Hamer@unsw.edu.au
\bibitem[\dag]{byline2}e-mail address: w.zheng@unsw.edu.au
\bibitem[\ddag]{byline3}e-mail address: otja@newt.phys.unsw.edu.au

\bibitem{dag94}E. Dagotto, Rev. Mod. Phys.  {\bf 66}, 763(1994).

\bibitem{dag96}E. Dagotto, T.M. Rice, Science  {\bf 271}, 618(1996);
	T.M. Rice, Z. Phys. B{\bf 103}, 165(1997).

\bibitem{Lu93}L. Yu, Z.B. Su, and Y.M. Li, Chin. J. Phys. {\bf 31}, 579(1993).

\bibitem{hir85}J. Hirsch, Phys. Rev. Lett. {\bf 54}, 1317(1985).
\bibitem{hyb90}M.S. Hybertson, E.B. Stechel, M. Schluter and
	D.R. Jennison, Phys. Rev. B {\bf 41}, 11068(1990).
\bibitem{zha88}F.C. Zhang and T.M. Rice, Phys. Rev. B {\bf 37}, 3759(1988).
\bibitem{man91}E. Manousakis, Rev. Mod. Phys. {\bf 63}, 1(1991).
\bibitem{bar89}T. Barnes, E. Dagotto, A. Moreo and
	E.S. Swanson, Phys. Rev. B{\bf 40}, 10977(1989);
	T. Barnes, A.E. Jacobs, M.D. Kovarik and
	W.G. Macready, Phys. Rev. B {\bf 45}, 256(1992).
\bibitem{poi94}D. Poilblanc, J. Riera and E. Dagotto, Phys.
Rev. B {\bf 49}, 12318(1994).

\bibitem{leu95}P.W. Leung and R.J. Gooding, Phys. Rev. B {\bf 52},
15711(1995).
\bibitem{lee97}T.K. Lee and C.T. Shih, Phys. Rev. B {\bf 55}, 5983(1997).
\bibitem{bon}M. Boninsegni and E. Manousakis, Phys. Rev. B 
{\bf 47}, 11897(1993).
\bibitem{liu92}Z.P. Liu and E. Manousakis, Phys. Rev. B {\bf 45}, 2425(1992).
\bibitem{ede92}R. Eder, Phys. Rev. B {\bf 45}, 319(1992).

\bibitem{yu95}M.Yu. Kuchiev and O.P. Sushkov, Phys. Rev. B {\bf 52}, 12977(1995).

\bibitem{li96}Y.M. Li, N. d'Ambrumenil, L. Yu and Z.B. Su,
Phys. Rev. B{\bf 53}, 14717(1996).

\bibitem{bel97}V.I. Belinicher, A.L. Chernyshev and V.A. Shubin,
Phys. Rev. B {\bf 56}, 3381(1997).

\bibitem{shr90}B. Shraiman and E. Siggia, Phys. Rev. Lett. {\bf 62}, 1564(1989).

\bibitem{kuc93}M. Kuchiev and O. Sushkov, Phys. Rev. B {\bf 52}, 12977(1995).

\bibitem{he90}H.X. He, C.J. Hamer and J. Oitmaa, J. Phys. A {\bf 23}, 1775(1990).

\bibitem{oit94}J. Oitmaa, C.J. Hamer and W.H. Zheng,
  Phys. Rev. B {\bf 50}, 3877(1994).

\bibitem{oit96}J. Oitmaa and W.H. Zheng, Phys. Rev. B {\bf 54}, 3022(1996).

\bibitem{sin89}R.R.P. Singh, Phys. Rev. B {\bf 39}, 9760(1989); {\bf 41}, 4873(1990).

\bibitem{gel90}
M. P. Gelfand, R. R. P. Singh and D. A. Huse,
J. Stat. Phys. {\bf 59}, 1093 (1990).

\bibitem{gelfand}
M. P. Gelfand,
Solid State Comm. {\bf 98}, 11 (1996).

\bibitem{shi95}Z.P. Shi and R.R.P. Singh, Phys. Rev. B{\bf 52}, 9620(1995);
Europhys. Lett. {\bf 3}, 219(1995).

\bibitem{put92}W.O. Putikka, M.U. Luchini and T.M. Rice,
   Phys. Rev. Lett. {\bf 68}, 538(1992); W.O. Putikka, M.U. Luchini and M. Ogato,
Phys. Rev. Lett. {\bf 69}, 2288(1992).

\bibitem{sin92}R.R.P. Singh and R.L. Glenister, Phys. Rev. B{\bf 46}, 11871(1992).

\bibitem{gut}A.J. Guttmann, in {\it Phase Transitions and Critical Phenomena},
edited by C. Domb and M.S. Green (Academic, New York, 1989), Vol. 13.

\bibitem{ede90}R. Eder and K.W. Becker and W.H. Stephan, Z. Phys. B {\bf 81}, 33 (1990);
R. Eder and K.W. Becker, Z. Phys. B {\bf 78}, 219 (1990).

\bibitem{mar91}G. Martinez and P. Horsch, Phys. Rev. B {\bf 44}, 317(1991).

\bibitem{bul89}N. Bulut, D. Hone and D.J. Scalapino, Phys. Rev. Lett.
{\bf 62}, 2192(1989).

\bibitem{bar90}T. Barnes and M.D. Kovarik, Phys. Rev. B {\bf 42}, 6159(1990).

\bibitem{sch89}S. Sachdev, Phys. Rev. B {\bf 39}, 12232(1989).

\bibitem{che98}A.L. Chernyshev, P.W. Leung and
R.J. Gooding, cond-mat/9806018.

\bibitem{bar92}M.A. Baranov and M.Y. Kagan, Z. Phys. B{\bf 86}, 237(1992);
M.A. Baranov, A.V. Chubukov and M.Y. Kagan, Int. J. Mod. Phys.
B{\bf 6}, 2411(1992).

\bibitem{chu93}A.V. Chubukov, Phys. Rev. Lett. B{\bf 48}, 1097(1993).

\bibitem{dag93}E. Dagotto and J. Riera, Phys. Rev. Lett. {\bf 70}, 682(1993);
E. Dagotto, J. Riera, Y.C. Chen, A. Moreo, A. Nazarenko,
F. Alcaraz and F. Ortolani, Phys. Rev. B{\bf 49}, 3548(1994).

\bibitem{hel97}C.S. Hellberg and E. Manousakis, Phys. Rev.
Letts. {\bf 78}, 4609(1997), and references therein.

\bibitem{kag94}M.Y. Kagan and T.M. Rice, J. Phys. Cond. Mat. {\bf 6}, 3771(1994).

\bibitem{rie93}J. Riera and E. Dagotto, Phys. Rev. B{\bf 47}, 15346(1993).

\bibitem{sca91}R.T. Scalettar, R.R.P. Singh and S.C. Zhang, 
Phys. Rev. Lett., {\bf 67}, 370(1991).


\bibitem{eme90}V.J. Emery, S.A. Kivelson and H.Q. Lin,
Phys. Rev. Lett. {\bf 64}, 475(1990); Phys. Rev. B{\bf 42}, 6523(1990).

\bibitem{poi95}D. Poiblanc, Phys. Rev. B {\bf 52}, 9201(1995).

\bibitem{koh97}M. Kohno, Phys. Rev. B {\bf 55}, 1435(1997).

\bibitem{shih98}C.T. Shih, Y.C. Chen, H.Q. Lin and T.K. Lee, cond-mat/9807027.

\end{references}
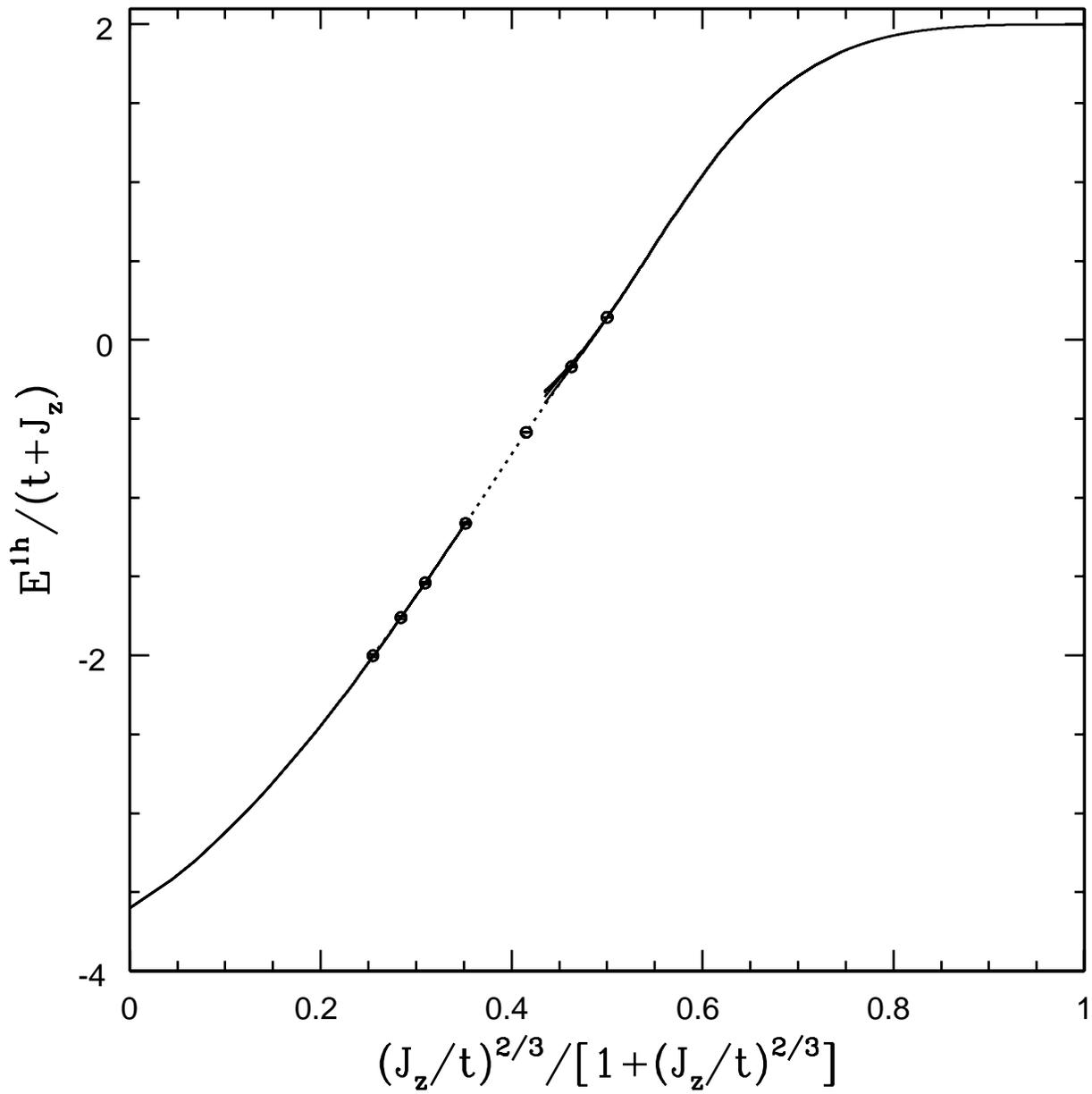
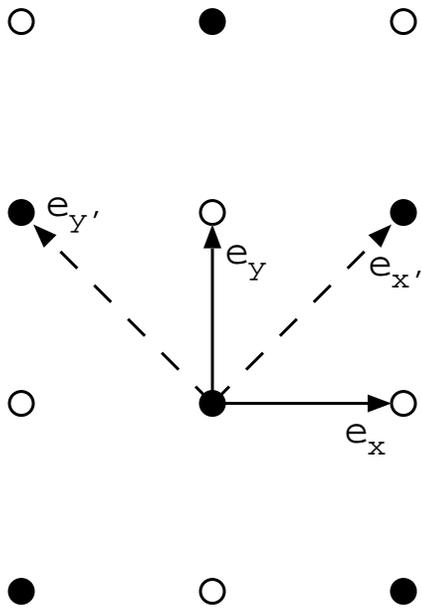
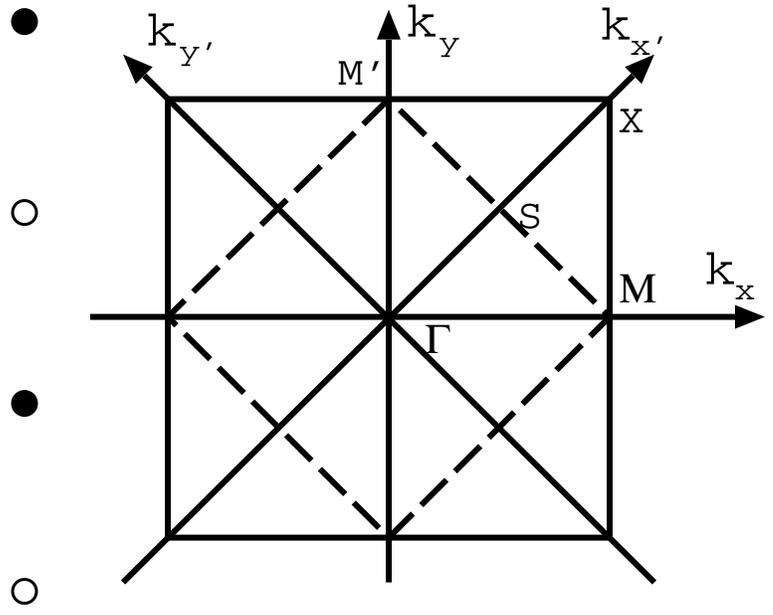
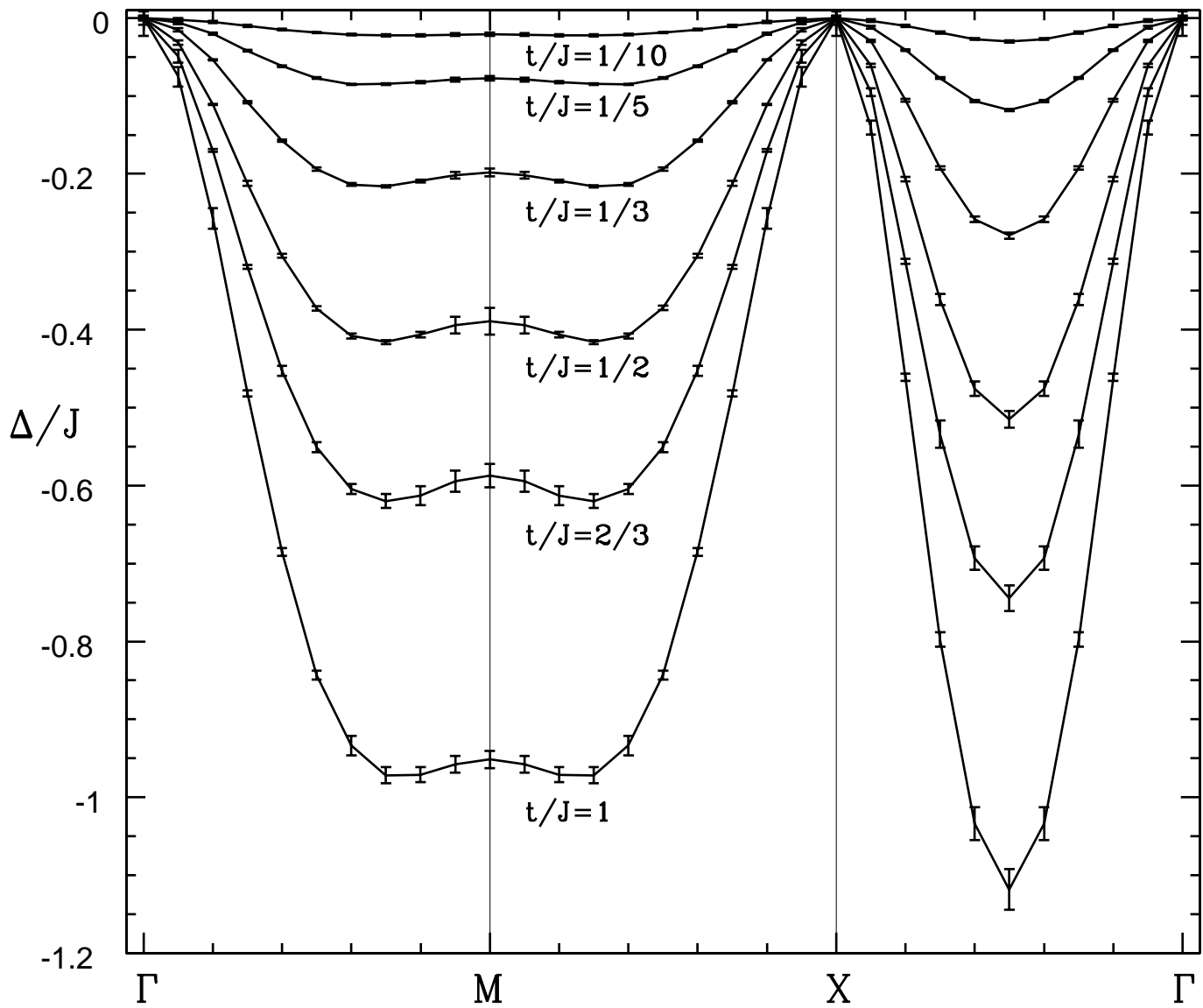
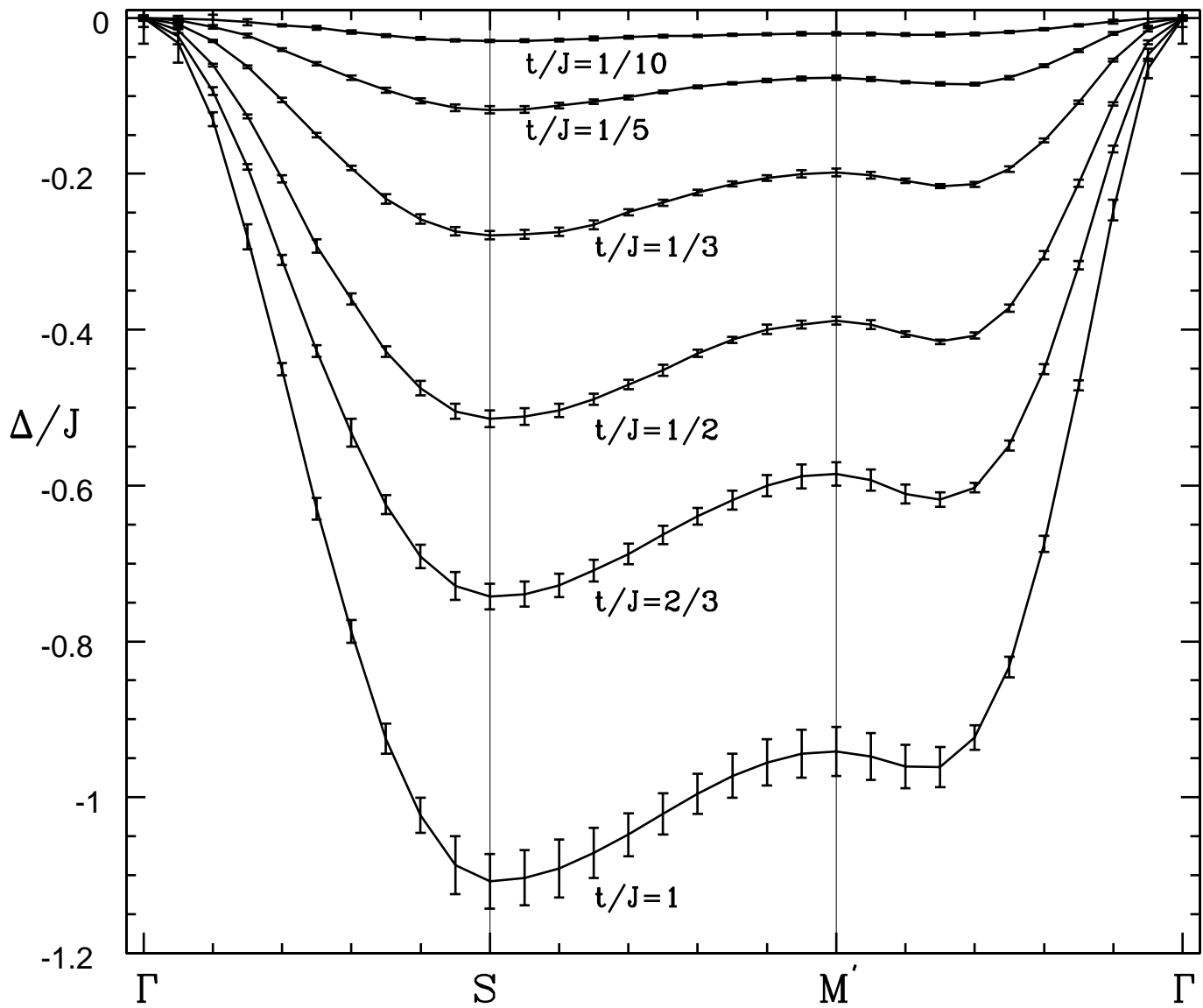
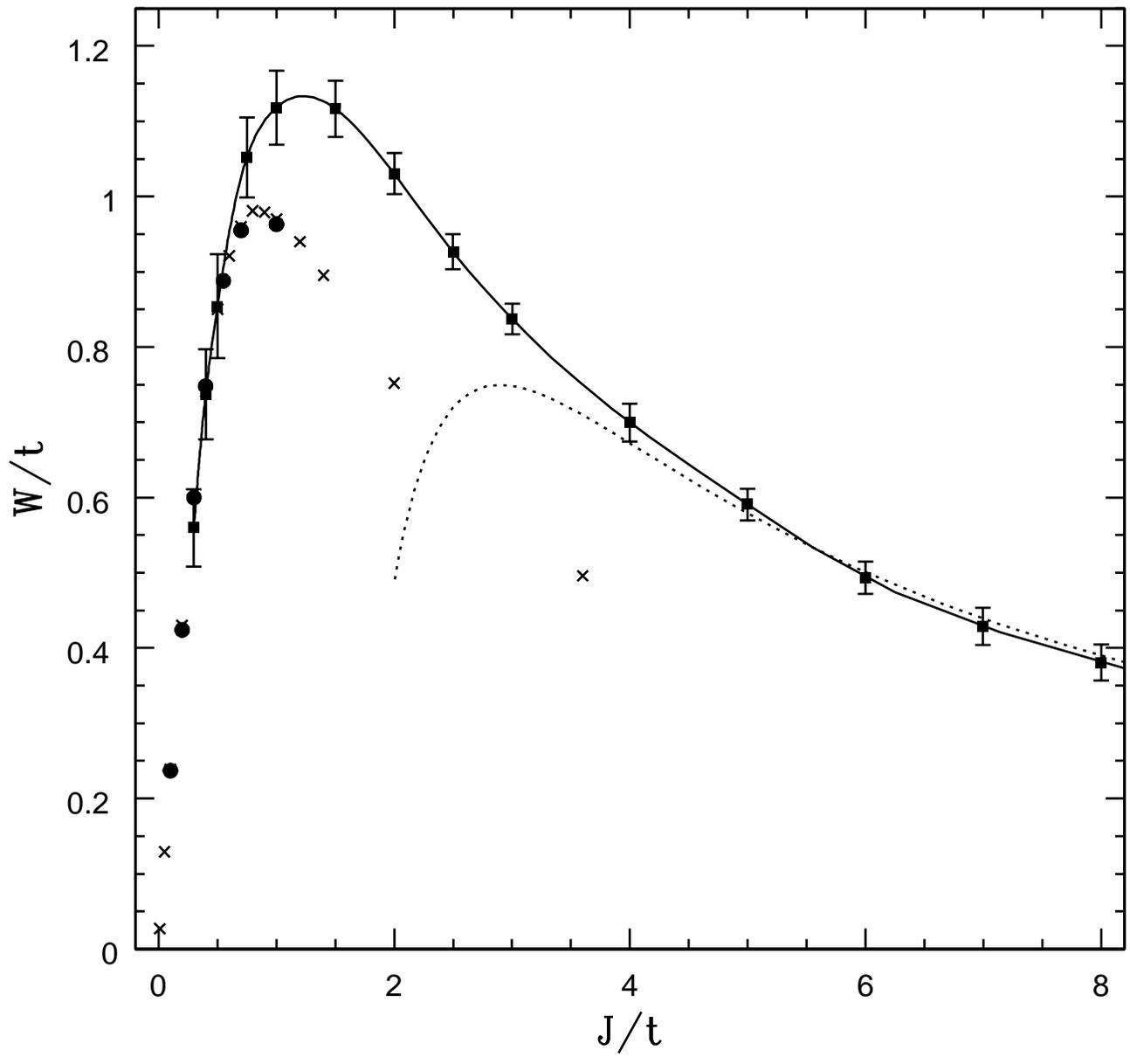
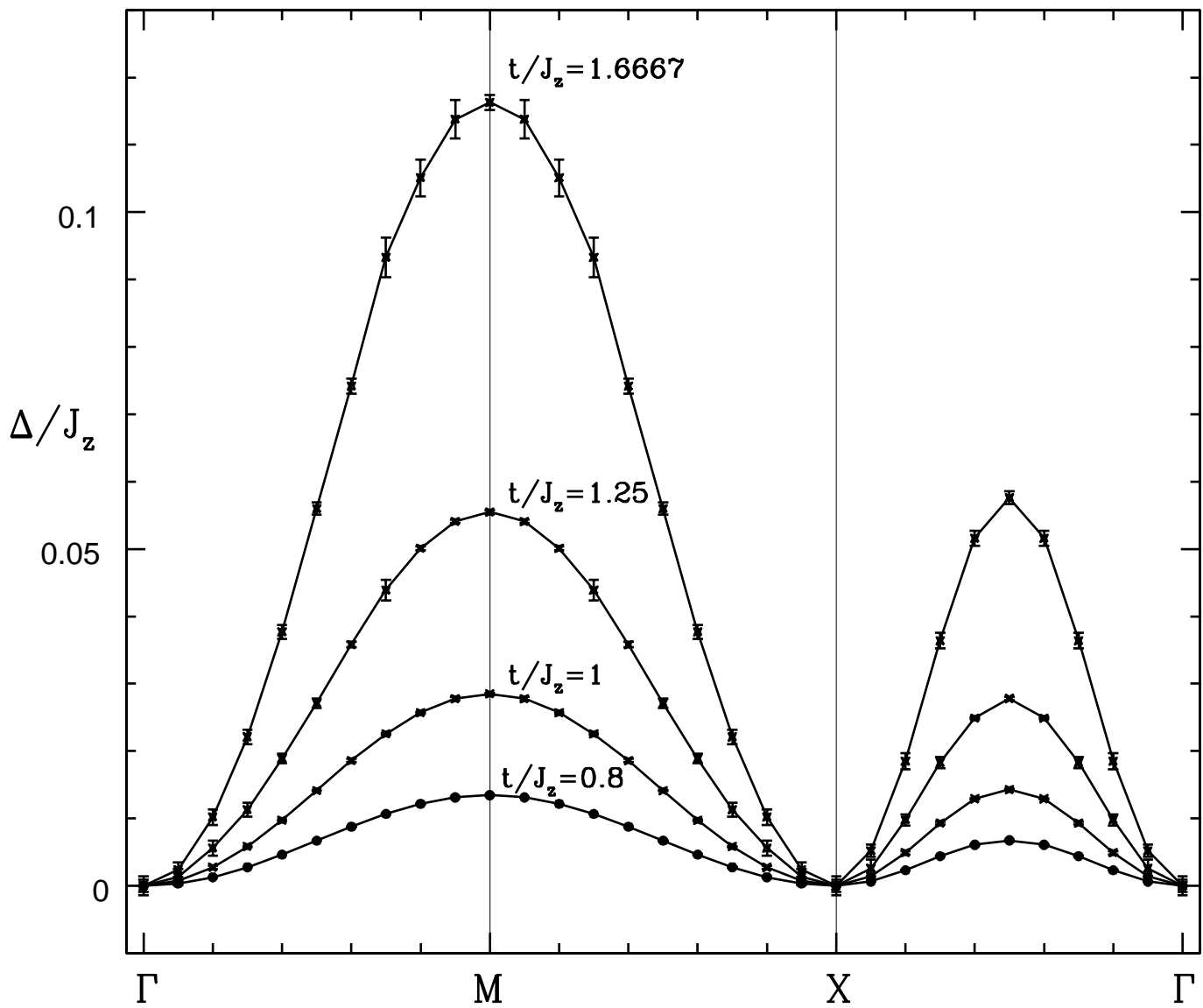
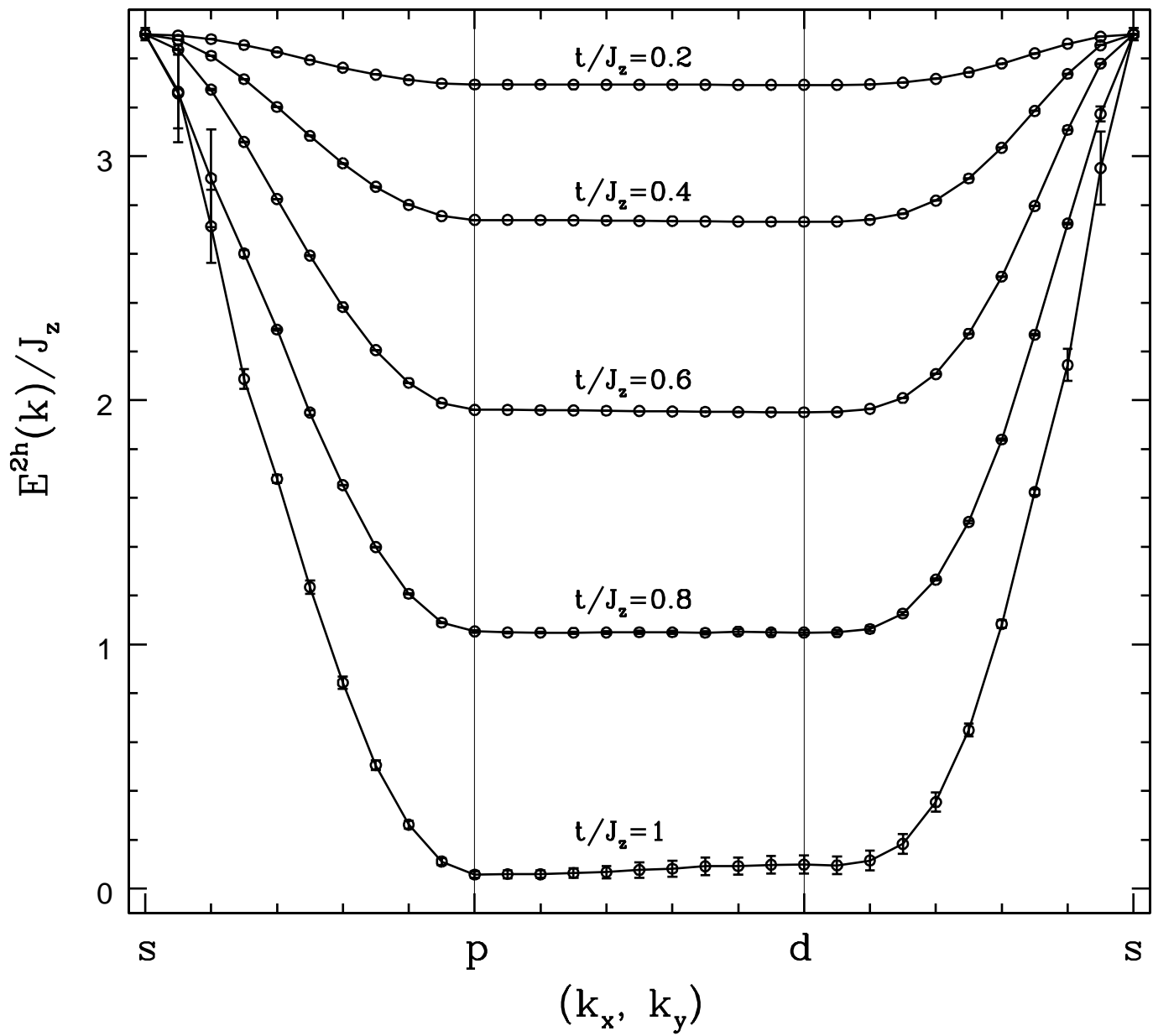
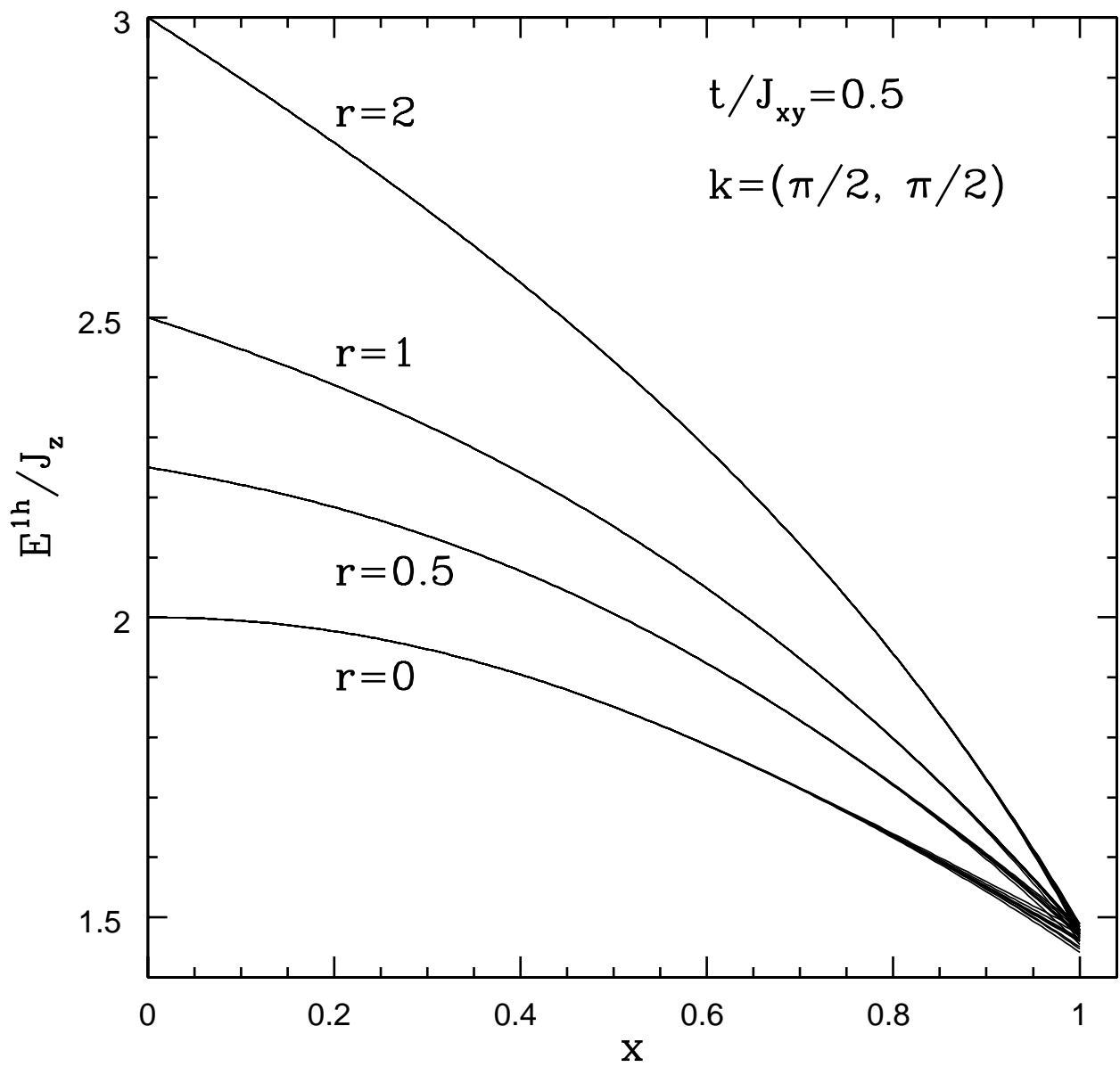
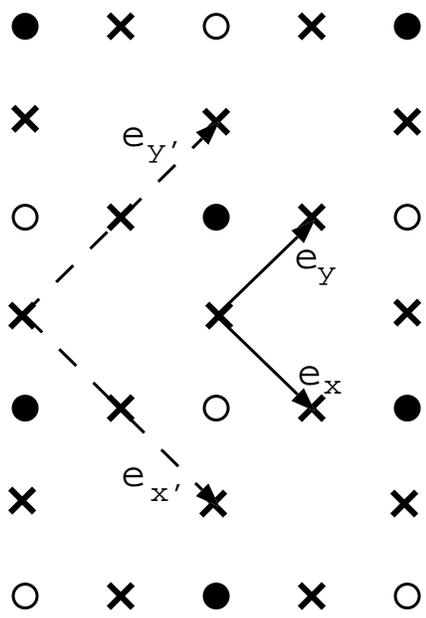
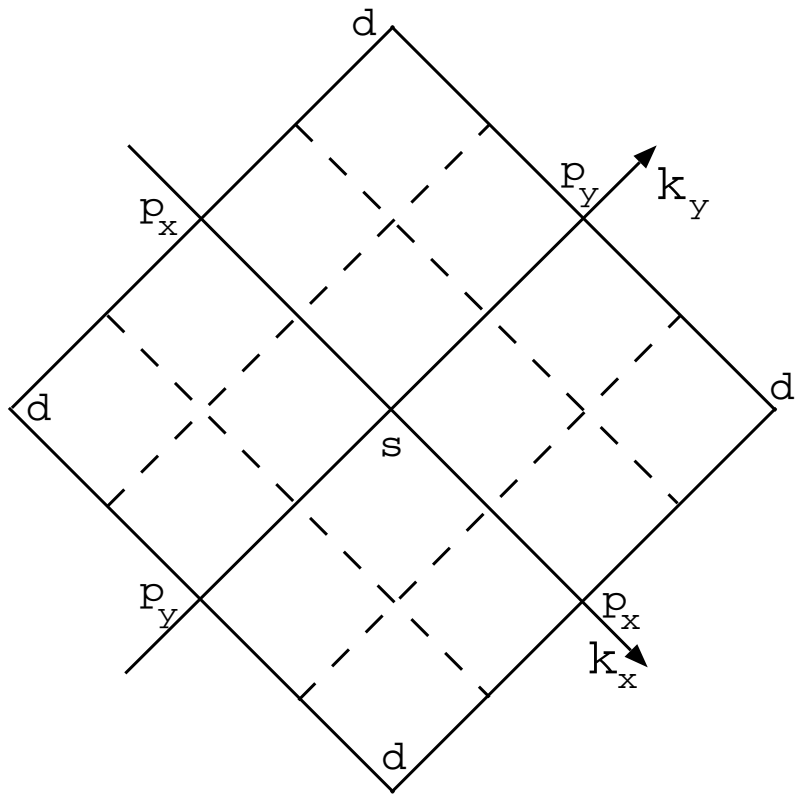
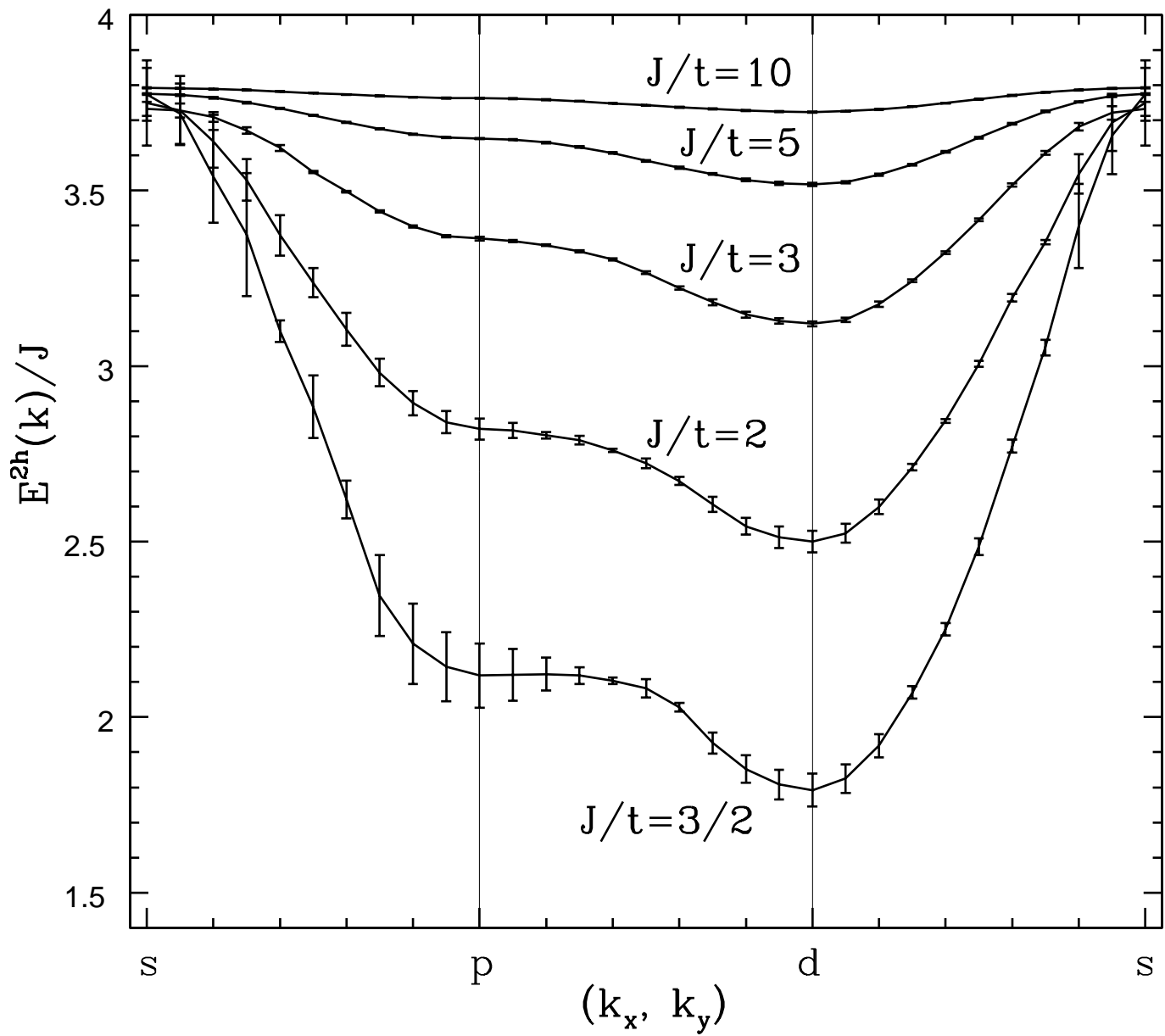
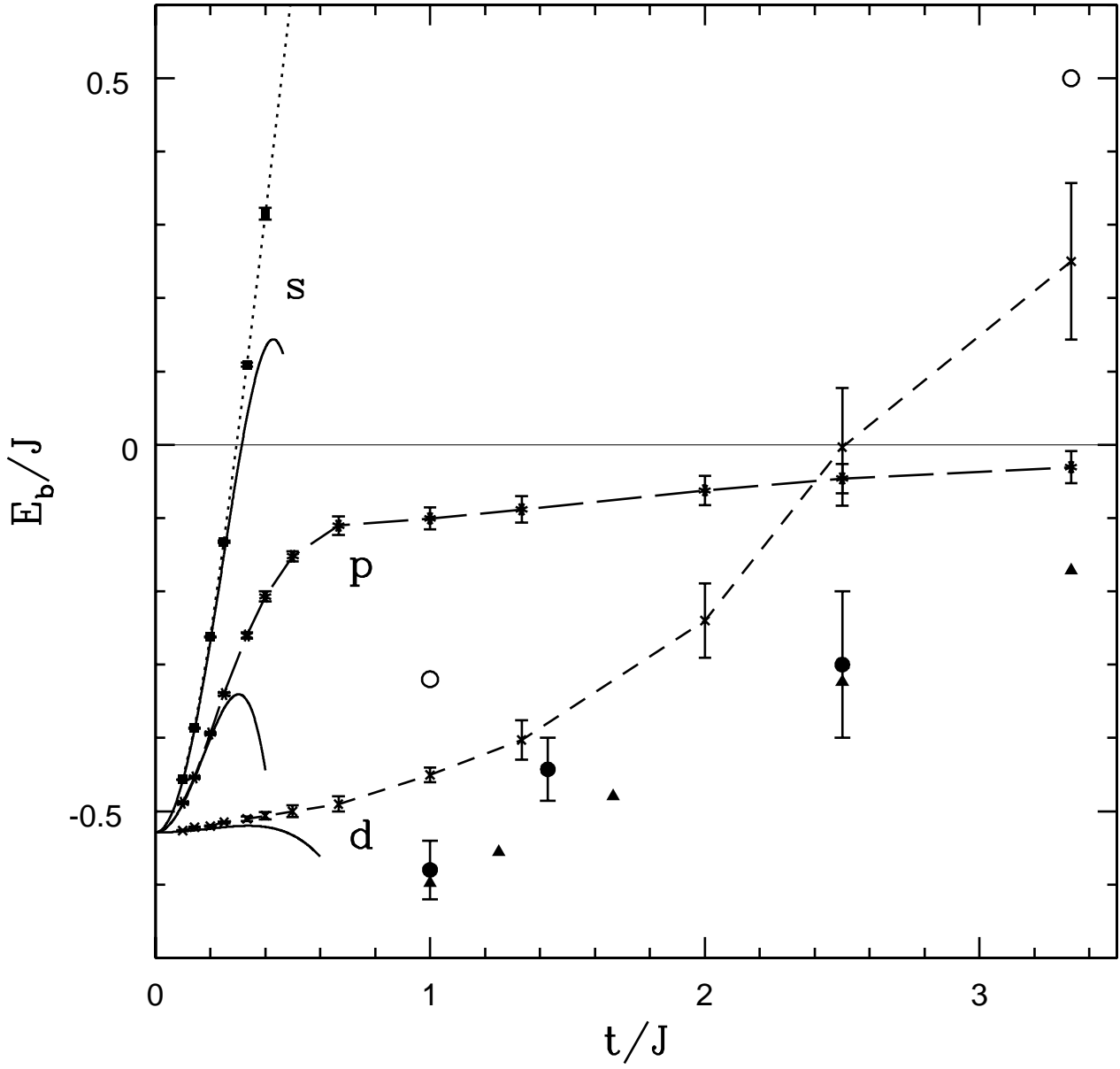
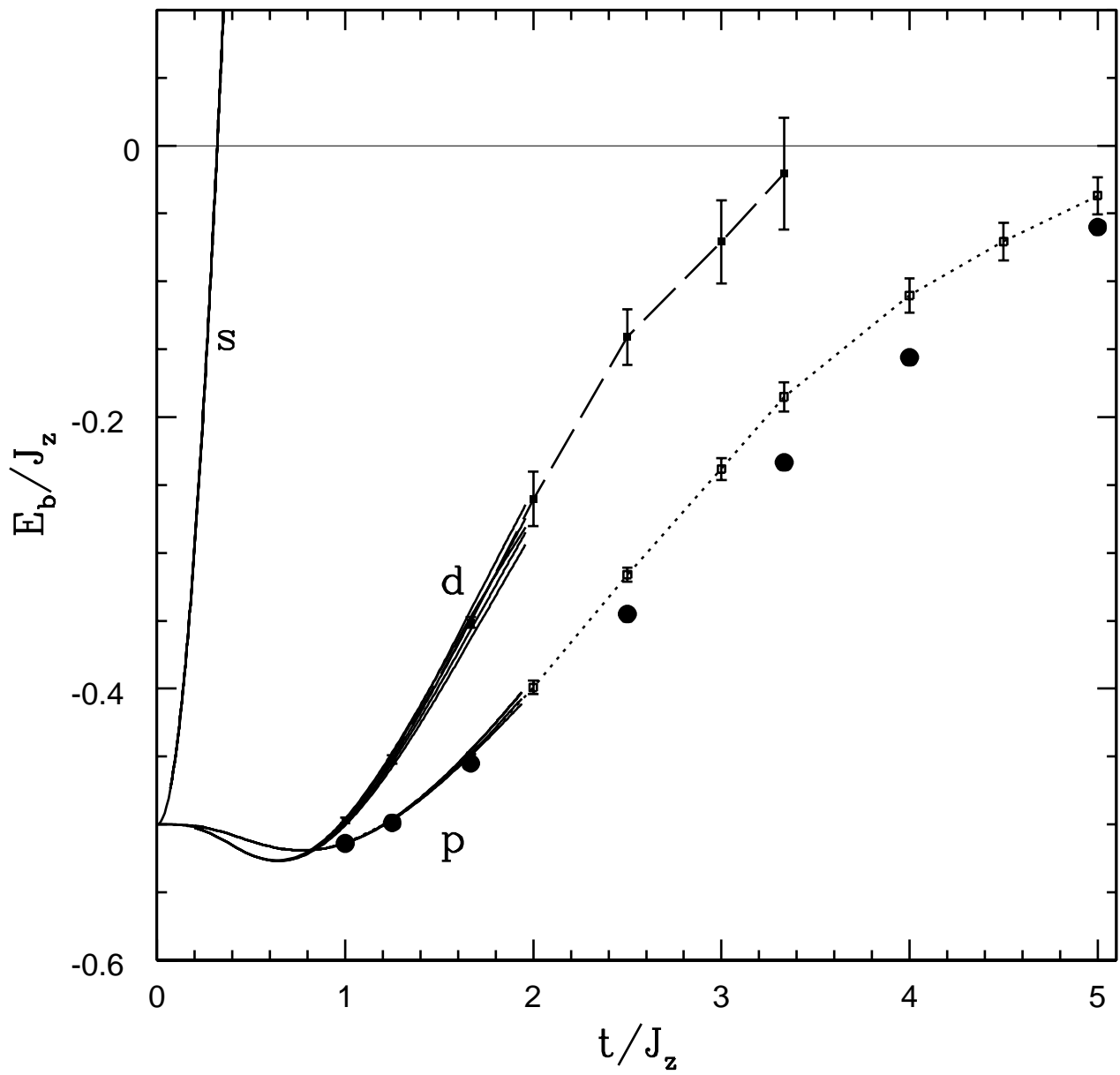
\end{document}